\documentclass[twocolumn,10pt]{revtex4}

\usepackage{graphicx}
\usepackage{dcolumn}
\usepackage{bm}

\input epsf

\begin{document}

\title{\Large Charged Gravastars in Rastall-Rainbow Gravity}

\author{\bf Ujjal Debnath\footnote{ujjaldebnath@gmail.com}}

\affiliation{Department of Mathematics, Indian Institute of
Engineering Science and Technology, Shibpur, Howrah-711 103,
India.\\}

\begin{abstract}
In this work, we have considered the spherically symmetric stellar
system in the contexts of Rastall-Rainbow gravity theory in
presence of isotropic fluid source with electro-magnetic field.
The Einstein-Maxwell's field equations have been written in the
framework of Rastall-Rainbow gravity. Next we have discussed the
geometry of charged gravastar model. The gravastar consists of
three regions: interior region, thin shell region and exterior
region. In the interior region, the gravastar follows the equation
of sate (EoS) $p=-\rho$ and we have found the solutions of all
physical quantities like energy density, pressure, electric field,
charge density, gravitational mass and metric coefficients. In the
exterior region, we have obtained the exterior Riessner-Nordstrom
solution for vacuum model ($p=\rho=0$). Since in the shell region,
the fluid source follows the EoS $p=\rho$ (ultra-stiff fluid) and
the thickness of the shell of the gravastar is infinitesimal, so
by the approximation $h~(\equiv A^{-1})\ll 1$, we have found the
analytical solutions within the thin shell. The physical
quantities like the proper length of the thin shell, entropy and
energy content inside the thin shell of the charged gravastar have
been computed and we have shown that they are directly
proportional to the proper thickness of the shell ($\epsilon$) due
to the approximation ($\epsilon\ll 1$). The physical parameters
significantly depend on the Rastall parameter and Rainbow
function. Next we have studied the matching between the surfaces
of interior and exterior regions of the charged gravastar and
using the matching conditions, the surface energy density and the
surface pressure have been obtained. Also the equation of state
parameter on the surface, mass of the thin shell, mass of the
gravastar have been obtained. Finally, we have explored the stable
regions of the charged gravastar in Rastall-Rainbow gravity.
\end{abstract}

\maketitle


\section{Introduction}

A {\it gravastar} is astronomically hypothetical condensed object
which is a gravitationally dark cold vacuum compact star or
gravitational vacuum condensate star. Mazur and Mottola
\cite{Mazur,Mazur1} have established the gravastar solution in the
concept of Bose-Einstein condensation to gravitational systems.
The gravastar is singularity free object which is spherically
symmetric as well as super compact. It has also the property that
it has no event horizon. So the gravastar is a substitute of black
hole i.e., the existence of compact stars minus event horizons.
The gravastar consists of three regions: (i) {\it Interior region}
($0 \le r < r_{1}$), (ii) {\it Shell region} ($r_{1} < r < r_{2}$)
and (iii) {\it Exterior region} ($r_{2}<r$), where $r_{1}$ and
$r_{2}$ are inner and outer radii ($r_{1}<r_{2}$). In the interior
region, the isotropic pressure produces a force of repulsion over
the intermediate thin shell. So the equation of state (EOS) of the
fluid satisfies $p=-\rho$ which describes the de-Sitter spacetime.
The intermediate thin shell region consists of ultra-stiff perfect
fluid satisfies the EoS $p=\rho$. The exterior region consists of
vacuum with EOS $p =\rho= 0$ which is described by the
Schwarzschild solution. Visser \cite{Visser} developed the
mathematical model of the gravastar and described the stability of
gravastar by taking some realistic values of EoS parameter.\\

DeBenedictis et al \cite{De} have found the gravastar solutions by
taking continuous pressures and the equation of state. The
anisotropic pressure for the structure of gravastar has been
considered by Cattoen et al \cite{Cattoen}. Bilic et al
\cite{Bilic} have found the gravastar solution in presence of
Born-infeld phantom model. Carter \cite{Carter} studied the
stability of the gravastar. The Gravastar solutions in the
framework of conformal motion have been investigated by some
authors \cite{Usmani,Bhar,Ayan}. Gravastar model in higher
dimensional spacetime has been discussed in refs
\cite{Bhar,Farook1,Ghos,Ghosh1}. Several authors
\cite{Bene,Rocha1,Rocha2,Chan1,R1,Lobo00,Ayan1} have discussed the
stable nature of gravastars. Charged gravastar models with its
physical features have been analyzed in the works
\cite{Usmani,Bhar,Ghos,Fel,Tur,Horvat,Chan,Rahaman1,Brandt,Yous00,Ali}.
Ray et al \cite{Ray} have described the charged strange quark star
model in the framework of electromagnetic mass with conformal
killing vector.\\

In the framework of modified gravity theory, lot of works on
compact star, neutron star, strange star and gravastar have been
found in the literature. Boehmer et al \cite{Boem} have examined
the existence of relativistic stars in $f(T)$ modified gravity.
The structure of neutron stars in modified $f(T)$ gravity has been
studied by Deliduman et al \cite{Del}. In refs \cite{Abha1,Saha1},
the authors have studied the anisotropic strange stars in $f(T)$
gravity model. The structures of relativistic stars in $f(T)$
gravity and its Tolman-Oppenheimer-Volkoff (TOV) equations have
been computed in \cite{Kp}. In $f(T)$ gravity model, the compact
star models have been studied in \cite{Abha2} and the neutron star
models have been discussed in \cite{Gan}. Using the Krori and
Barua (KB) metric \cite{Krori}, the anisotropic compact star
models in GR, $f(R)$, $f(G)$ and $f(T)$ theories have been studied
in refs \cite{Abbas6,Abbas7,Abbas8,Abbas9}. The gravastar solution
in $f(R,{\cal T})$ gravity model has been studied in
\cite{Das1,Yous}. Also the Gravastar solution in $f(G,{\cal T})$
gravity model has been found in \cite{Shamir}.\\

Rastall gravity theory is proposed by Rastall \cite{Rastall},
which is one of the alternative of modified gravity theory by
modification the Einstein's general relativity. Neutron Stars in
Rastall Gravity have been obtained by Oliveira et al \cite{Oliv}.
A model of quintessence compact stars in the Rastall's theory of
gravity has been obtained by Abhas et al \cite{AbhasS}. Isotropic
compact star model in Rastall theory admitting conformal motion
has also been obtained in \cite{AbhasS1}. Anisotropic compact star
model in the Rastall theory of gravity has been discussed in
\cite{AbhasS2}. Gravity's rainbow \cite{Mag} is a distortion of
space-time which is an extension of the doubly special relativity
for curved space-times. The properties of neutron stars and
dynamical stability conditions in the modified TOV in gravity's
rainbow have been investigated by Hendi et al \cite{Hen}. The
gravity's rainbow and compact star models have been studied in
\cite{Gar}. The rainbow's star models have also been studied in
\cite{Gara}. Recently, Mota et al \cite{Mota0} have studied the
neutron star model in the framework of Rastall-Rainbow theories of gravity.\\

The main motivation of the work is to study the gravastar system
in the framework of Rastall-Rainbow gravity with the isotropic
fluid and electromagnetic source and examine the nature of
physical parameters and stability of the gravastar. The
organization of the work is as follows: In section II, we present
the Rastall-Rainbow gravity theory with electromagnetic field.
Here we write the Einstein-Maxwell field equations for spherically
symmetric steller metric in the framework of Rastall-Rainbow
gravity. Section III deals with the geometry of gravastar and we
compute the solutions in the three regions of gravastar model. In
section IV, we analyze the physical aspects of the parameters of
gravastar model. In section V, we investigate the matching between
interior and exterior regions. Due to the junction conditions, we
compute the equation of state, mass of the thin shell region and
examine the stability of the gravastar. Finally some physical
analysis and fruitful conclusions of the work are drawn in section VI.\\

\section{Rastall-Rainbow Gravity}

In Einstein's General Relativity (GR), the conservation law of
energy-momentum tensor is $T_{\mu;\nu}^{\nu}=0$. The Rastall's
Gravity is a generalization of General Relativity, where Rastall
\cite{Rastall} proposed the modification of conservation law of
energy-momentum tensor in curved space-time and which is given by
\cite{Mota0}
\begin{equation}\label{1}
T_{\mu;\nu}^{\nu}=\bar{\lambda}R_{,\mu}
\end{equation}
where $\bar{\lambda}=\frac{1-\lambda}{16\pi G}$ with $\lambda$ is
a constant called Rastall parameter, which measures the deviation
from GR and describes the affinity of the matter field to couple
with geometry. For $\lambda=1$, the usual conservation law can be
restored. Also for flat space-time, the Ricci scalar $R=0$ and we
may also the usual conservation law. So for the effect of
Rastall's gravity, $\lambda\ne 1$ and the space-time must be
non-flat. The above equation can be written as
\begin{equation}
\left(T_{\mu}^{\nu}-\bar{\lambda}\delta_{\mu}^{\nu}R\right)_{;\nu}=0
\end{equation}
So for Rastall's gravity, the Einstein's equation can be modified
to the form
\begin{equation}
R_{\mu}^{\nu}-\frac{1}{2}\delta_{\mu}^{\nu}R=8\pi G
\left(T_{\mu}^{\nu}-\bar{\lambda}\delta_{\mu}^{\nu}R\right)
\end{equation}
which can be simplified to the form
\begin{equation}
R_{\mu}^{\nu}-\frac{\lambda}{2}\delta_{\mu}^{\nu}R=8\pi G
T_{\mu}^{\nu}
\end{equation}
Now the trace of the energy-momentum tensor is given by
\begin{equation}
T=\frac{(1-2\lambda)R}{8\pi G}
\end{equation}
So the above Einstein's equation in Rastall's gravity can be
written as
\begin{equation}
R_{\mu}^{\nu}-\frac{1}{2}\delta_{\mu}^{\nu}R=8\pi G
\left(T_{\mu}^{\nu}-\frac{(1-\lambda)}{2(1-2\lambda)}\delta_{\mu}^{\nu}T\right)
\end{equation}
Now assume that the fluid source is composed of normal matter and
electro-magnetic field. So the energy momentum tensor can be
written as
\begin{equation}
T_{\mu\nu}=T_{\mu\nu}^{M}+T_{\mu\nu}^{EM}
\end{equation}
where the energy-momentum tensor for normal matter is given by
\begin{equation}
T_{\mu\nu}^{M}=(\rho+p)u_{\mu}u_{\nu}+pg_{\mu\nu}
\end{equation}
where $u^{\mu}$ is the fluid four-velocity satisfying
$u_{\mu}u^{\nu}=-1$, $\rho$ and $p$ are the energy density and
pressure of fluid. Further, the energy momentum tensor for
electromagnetic field is given by \cite{Debn}
\begin{equation}\label{7}
T_{\mu\nu}^{EM}=-\frac{1}{4\pi}(g^{\delta\omega}F_{\mu\delta}F_{\omega\nu}
-\frac{1}{4}g_{\mu\nu}F_{\delta\omega}F^{\delta\omega})
\end{equation}
where $F_{\mu\nu}$ is the Maxwell field tensor defined as in the
form:
\begin{equation}\label{8}
F_{\mu\nu}=\Phi_{\nu,\mu}-\Phi_{\mu,\nu}
\end{equation}
and $\Phi_{\mu}$ is the four potential. The corresponding
equations for Maxwell's electromagnetic field are given by
\begin{equation}
(\sqrt{-g}~F^{\mu\nu}),_{\nu}=4\pi
J^{\mu}\sqrt{-g}~,~F_{[\mu\nu,\delta]}=0
\end{equation}
where $J^{\mu}$ is the current four-vector satisfying
$J^{\mu}=\sigma u^{\mu}$, the parameter $\sigma$ is the charge
density.\\

Magueijo and Smolin \cite{Mag} have proposed gravity's Rainbow,
which is an extension of the doubly special relativity for curved
space-times. Gravity's rainbow is a distortion of space-time
induced by two arbitrary functions $\Pi(x)$ and $\Sigma(x)$
(called the Rainbow functions) satisfying
\begin{equation}
{\cal E}^{2}\Pi^{2}(x)-\upsilon^{2}\Sigma^{2}(x)=m^{2}
\end{equation}
where $x={\cal E}/{\cal E}_{Pl}$. Here ${\cal E}$, $\upsilon$, $m$
and ${\cal E}_{Pl}=\sqrt{\hbar c^{5}}/G$ are the energy, momentum,
mass of a test particle and Planck energy respectively. Awad et al
\cite{Awad} and Khodadi et al \cite{Khod} have chosen $\Pi(x)=1$
and $\Sigma(x)=\sqrt{1+x^{2}}$ to study the solutions
corresponding to a nonsingular universe. Also to study of gamma
ray burst, the exponential form of rainbow has been applied in
\cite{Awad,Am}. In the absence of the test particles, the Rainbow
functions are satisfying
\begin{equation}
\lim_{x \to 0}\Pi(x)=1,~\lim_{x \to 0}\Sigma(x)=1
\end{equation}

Mota et al \cite{Mota0} have merged the Rastall's gravity with
Rainbow's gravity and applied the Rastall-Rainbow gravity theory
in the neutron star formation. We consider the spherically
symmetric metric describing the interior space-time of a star in
Rainbow gravity as \cite{Mota0}
\begin{equation}\label{9}
ds^{2}=-\frac{B(r)}{\Pi^{2}(x)}dt^{2}+\frac{A(r)}{\Sigma^{2}(x)}dr^{2}
+\frac{r^{2}}{\Sigma^{2}(x)}(d\theta^{2}+\sin^{2}\theta d\phi^{2})
\end{equation}
where $A(r)$ and $B(r)$ are functions of $r$. Since $\Pi(x)$ and
$\Sigma(x)$ depend on $x={\cal E}/{\cal E}_{Pl}$ and ${\cal E}$ is
independent of $r$, so $\Pi(x)$ and $\Sigma(x)$ are independent of
$r$. The metric coefficients depend of the energy of the test
particle. So the geometry of the space-time becomes energy
dependent. It should be noted that the metric coordinates do not
depend on the energy of the particle.\\

For the charged fluid source with density $\rho(r)$, pressure
$p(r)$ and electromagnetic field $E(r)$, the Einstein-Maxwell (EM)
equations in the Rastall-Rainbow gravity can be written in the
form \cite{Mota0}
\begin{eqnarray}\label{1st}
\frac{A'}{rA^{2}}-\frac{1}{r^{2}A}+\frac{1}{r^{2}}=8\pi
G\bar{\rho}~,
\end{eqnarray}
\begin{equation}\label{2nd}
\frac{B'}{rAB}+\frac{1}{r^{2}A}-\frac{1}{r^{2}}=8\pi
G\bar{p}_{1}~,
\end{equation}
and
\begin{eqnarray}\label{3rd}
\frac{B''}{2AB}-\frac{A'B'}{4A^{2}B}-\frac{B'^{2}}{4AB^{2}}-\frac{A'}{2rA^{2}}+\frac{B'}{2rAB}=8\pi
G\bar{p}_{2}
\end{eqnarray}
where
\begin{eqnarray}
\bar{\rho}=\frac{1}{\Sigma^{2}(x)}\left(\alpha_{1}\rho+3\alpha_{2}p+\frac{1}{8\pi}(\alpha_{1}-3\alpha_{2})E^{2}
\right)~,
\end{eqnarray}
\begin{eqnarray}
\bar{p}_{1}=\frac{1}{\Sigma^{2}(x)}\left(\alpha_{2}\rho+(1-3\alpha_{2})p+\frac{1}{8\pi}(4\alpha_{2}-1)E^{2}
\right)~,
\end{eqnarray}
and
\begin{eqnarray}
\bar{p}_{2}=\frac{1}{\Sigma^{2}(x)}\left(\alpha_{2}\rho+(1-3\alpha_{2})p+\frac{1}{8\pi}(1-2\alpha_{2})E^{2}
\right)
\end{eqnarray}
with
\begin{eqnarray}
\alpha_{1}=\frac{3\lambda-1}{2(2\lambda-1)}~,~\alpha_{2}=\frac{\lambda-1}{2(2\lambda-1)}
\end{eqnarray}
We observe that $\alpha_{1}+\alpha_{2}=1$ and $\lambda\ne 1/2$.
Adding equations (\ref{1st}) and (\ref{2nd}), we obtain
\begin{equation}\label{4th}
\frac{1}{rA}\left(\frac{A'}{A}+\frac{B'}{B} \right)=8\pi
G(\bar{\rho}+\bar{p}_{1})=\frac{8\pi G}{\Sigma^{2}(x)}(\rho+p)
\end{equation}
From equation (\ref{1st}), we obtain \cite{Mota0}
\begin{equation}\label{5th}
A(r)=\left(1-\frac{2GM(r)}{r} \right)^{-1}
\end{equation}
where the gravitational mass is
\begin{equation}\label{6th}
M(r)=\int_{0}^{r}4\pi r^{2}\bar{\rho}(r)dr
\end{equation}
From the modification of conservation law of energy-momentum
tensor, we obtain
\begin{equation}\label{7th}
p'+(\rho+p)\frac{B'}{2B}=\frac{1}{8\pi r^{4}}(r^{4}E^{2})'
\end{equation}
and the electric field $E$ is as follows
\begin{equation}\label{8th}
E(r)=\frac{1}{r^{2}}\int_{0}^{r}4\pi r^{2}\sigma(r)
\frac{\sqrt{A(r)}}{\Sigma(x)}~dr
\end{equation}
The term $\frac{\sigma \sqrt{A}}{\Sigma(x)}$ inside the above
integral is known as the volume charge density. \\

\section{Gravastar}

Here, we will derive the solutions of the field equations for
charged gravastar (charge is generated by electro-magnetic field)
and analyze its physical as well as geometrical interpretations.
Since there are three regions of the gravastar, so the geometrical
regions of the gravastar having a finite extremely thin width
within the regions $D=r_{1}<r<r_{2}=D+\epsilon$ where $r_{1}$ and
$r_{2}$ are radii of the interior region and exterior region of
the gravastar and $\epsilon$ is very small positive quantity. The
three regions are structured as follows: (i) {\it Interior region}
$R_{1}$: $0 \le r < r_{1}$ with the equation of state (EOS)
follows $p=-\rho$, (ii) {\it Shell region} $R_{2}$: $r_{1} < r <
r_{2}$ with EOS follows $p=\rho$ and (iii) {\it Exterior region}
$R_{3}$: $r_{2}<r$ with EOS follows $p =\rho= 0$.

\subsection{Interior Region}

The interior region $R_{1}$  ($0\le r<r_{1}=D$) of the gravastar
follows the EoS $p=-\rho$. From equation (\ref{3rd}), we obtain
the relation
\begin{equation}
B(r)=kA^{-1}(r)
\end{equation}
where $k$ is constant $>0$. There are 4 equations and 5 unknown
functions $A,~B,~\rho,~p,~E$ in the system. So one function is
free to us. Here we may consider $E$ is free function of $r$ and
is chosen by
\begin{equation}
E(r)=E_{0}r^{m}
\end{equation}
where $m$ is positive constant and $E_{0}$ is function of $x$.\\

Using equation (\ref{7th}), we obtain
\begin{equation}
p=-\rho=k_{1}r^{2m}-k_{2}
\end{equation}
where $k_{1}=\frac{(m+2)E_{0}^{2}}{8\pi m}$ and $k_{2}$ is
positive constant. From (\ref{6th}), we obtain
\begin{equation}
M(r)=\frac{1}{2(2\lambda-1)\Sigma^{2}(x)}\left(\frac{8\pi
k_{2}}{3}~r^{3}-\frac{2E_{0}^{2}}{m(2m+3)}~r^{2m+3} \right)
\end{equation}
From equation (\ref{5th}), we obtain the solution
\begin{eqnarray}
B(r)=kA^{-1}(r)=k\left[1-\frac{8\pi Gk_{2}}{3(2\lambda-1)\Sigma^{2}(x)}~r^{2} \right. \nonumber \\
\left. +\frac{2GE_{0}^{2}}{m(2m+3)(2\lambda-1)\Sigma^{2}(x)}
~r^{2m+2} \right]
\end{eqnarray}
So the metric becomes (choose $k=1$)
\begin{eqnarray}
&&ds^{2}=-\frac{1}{\Pi^{2}(x)}\left[1-\frac{8\pi
Gk_{2}}{3(2\lambda-1)\Sigma^{2}(x)}~r^{2}\right. \nonumber\\
&&\left.+\frac{2GE_{0}^{2}}{m(2m+3)(2\lambda-1)\Sigma^{2}(x)}
~r^{2m+2} \right]
  dt^{2}  \nonumber\\
&&  +\frac{1}{\Sigma^{2}(x)} \left[1-\frac{8\pi
  Gk_{2}}{3(2\lambda-1)\Sigma^{2}(x)}~r^{2}\right.\nonumber\\
&& \left. +\frac{2GE_{0}^{2}}{m(2m+3)(2\lambda-1)\Sigma^{2}(x)}
~r^{2m+2} \right]^{-1}
  dr^{2} \nonumber\\
&&  +\frac{r^{2}}{\Sigma^{2}(x)}(d\theta^{2}+\sin^{2}\theta
d\phi^{2})
\end{eqnarray}
The charge density for electric field is obtained in the form:
\begin{eqnarray}
&&\sigma(r)=\frac{(m+2)E_{0}r^{m-1}}{4\pi}\left[1-\frac{8\pi
Gk_{2}}{3(2\lambda-1)\Sigma^{2}(x)}~r^{2}\right.\nonumber\\
&&\left.+\frac{2GE_{0}^{2}}{m(2m+3)(2\lambda-1)\Sigma^{2}(x)}
~r^{2m+2} \right]^{\frac{1}{2}}
\end{eqnarray}
Also the gravitational mass of the interior region of the charged
gravastar can be found as
\begin{eqnarray}
M(D)=\frac{1}{2(2\lambda-1)\Sigma^{2}(x)}\left(\frac{8\pi
k_{2}}{3}~D^{3}-\frac{2E_{0}^{2}}{m(2m+3)}~D^{2m+3}
\right)\nonumber\\
\end{eqnarray}
We observe that the quantities $A(r),~B(r),~M(r),~\sigma(r)$
depend on the Rastall parameter $\lambda$ and Rainbow function
$\Sigma(x)$.

\subsection{Shell Region}
In the shell region $R_{2}$ ($D=r_{1}<r<r_{2}=D+\epsilon$), we
assume that the thin shell region contains stiff perfect fluid
which obeys EoS $p=\rho$. For this EoS, it is very difficult to
obtain the solution from the field equations. So we shall assume
the limit $0<A^{-1}\equiv h\ll 1$ in the thin shell to obtain the
analytical solution within the thin shell. Under this
approximation (we can set $h\approx 0$ to the leading order) with
$p=\rho$, the field equations (\ref{1st}) - (\ref{3rd}) reduce to
the following forms
\begin{equation}
-\frac{h'}{r}+\frac{1}{r^{2}}=\frac{8\pi
G}{\Sigma^{2}(x)}~\left[(\alpha_{1}+3\alpha_{2})\rho+\frac{1}{8\pi}~(\alpha_{1}-3\alpha_{2})E^{2}
\right]~,
\end{equation}
\begin{equation}\label{eq}
-\frac{1}{r^{2}}=\frac{8\pi
G}{\Sigma^{2}(x)}~\left[(1-2\alpha_{2})\rho+\frac{1}{8\pi}~(4\alpha_{2}-1)E^{2}
\right]
\end{equation}
and
\begin{equation}
\left(\frac{B'}{4B}+\frac{1}{2r}\right)h^{'}=\frac{8\pi
G}{\Sigma^{2}(x)}~\left[(1-2\alpha_{2})\rho+\frac{1}{8\pi}~(1-2\alpha_{2})E^{2}\right]
\end{equation}
Eliminating $\rho$ from the above equations, we ultimately get
following two equations
\begin{equation}\label{h1}
-\frac{\lambda
h'}{r}+\frac{2(2\lambda-1)}{r^{2}}=\frac{2G}{\Sigma^{2}(x)}~E^{2}
\end{equation}
and
\begin{equation}\label{h2}
\left(\frac{B'}{4B}+\frac{1}{2r}\right)h^{'}+\frac{1}{r^{2}}=\frac{(\lambda+1)G}{(2\lambda-1)\Sigma^{2}(x)}~E^{2}
\end{equation}
We see that there are two equations but three unknowns $B,~h$ and
$E$. Similar to the assumption of interior region, let us assume
that the solution of $E$ is in the form $E(r)=E_{0}r^{m}$. Solving
equations (\ref{h1}) and (\ref{h2}), we obtain
\begin{equation}\label{55}
A^{-1}(r)\equiv
h(r)=h_{1}+\frac{2(2\lambda-1)}{\lambda}~logr-\frac{E_{0}^{2}G}{\lambda(m+1)\Sigma^{2}(x)}~r^{2m+2}
\end{equation}
and
\begin{equation}
B(r)=\frac{h_{2}}{r^{2}}~Exp\left[\int\frac{4\lambda[(\lambda+1)E_{0}^{2}Gr^{2m+2}-(2\lambda-1)\Sigma^{2}(x)]~dr}
{r(2\lambda-1)[2(2\lambda-1)\Sigma^{2}(x)-2E_{0}^{2}Gr^{2m+2}]}
\right]
\end{equation}
where $h_{1}$ and $h_{2}$ are integration constants. In this shell
region $R_{2}$, the range of $r$ is satisfying $D< r< D+\epsilon$.
Since $h\ll 1$, so $\epsilon\ll 1$. Under this assumption, we must
have $h_{1}\ll 1$. From equation (\ref{eq}) we obtain
\begin{equation}
p=\rho=\frac{E_{0}^{2}r^{2m}}{8\pi\lambda}+\frac{(1-2\lambda)\Sigma^{2}(x)}{8\pi
G\lambda r^{2}}
\end{equation}
Also the charge density for electric field can be obtained in the
form
\begin{eqnarray}
&&\sigma(r)=\frac{(m+2)E_{0}r^{m-1}}{4\pi}\left[h_{1}+\frac{2(2\lambda-1)}{\lambda}~logr
\right.
\nonumber\\
&&\left. -\frac{E_{0}^{2}G}{\lambda(m+1)\Sigma^{2}(x)}~r^{2m+2}
\right]^{\frac{1}{2}}
\end{eqnarray}
We observe that the quantities $A(r),~B(r),~\sigma(r),~p,~\rho$
depend on the Rastall parameter $\lambda$ and Rainbow function
$\Sigma(x)$.

\subsection{Exterior Region}
The exterior region $R_{3}$ ($r>r_{2}=D+\epsilon$) contains the
vacuum whose EoS is given by $p=\rho=0$. In this exterior region,
using equation (\ref{7th}), we obtain
\begin{equation}
E(r)=\frac{Q}{r^{2}}
\end{equation}
where $Q$ is constant electric charge. We obtain the solutions as
\begin{equation}
B(r)=kA^{-1}(r)=k\left(1-\frac{2GM}{
r}+\frac{GQ^{2}}{(2\lambda-1)\Sigma^{2}(x) r^{2}}\right)
\end{equation}
where $M$ is the mass of the charged gravastar.\\

So in the exterior region, the metric becomes (choose $k=1$)
\begin{eqnarray}
&&ds^{2}=-\frac{1}{\Pi^{2}(x)}\left(1-\frac{2GM}{
r}+\frac{GQ^{2}}{(2\lambda-1)\Sigma^{2}(x) r^{2}}\right)
  dt^{2}  \nonumber\\
&&  +\frac{1}{\Sigma^{2}(x)} \left(1-\frac{2GM}{
r}+\frac{GQ^{2}}{(2\lambda-1)\Sigma^{2}(x) r^{2}}\right)^{-1}
  dr^{2}\nonumber\\
&&  +\frac{r^{2}}{\Sigma^{2}(x)}(d\theta^{2}+\sin^{2}\theta
d\phi^{2})
\end{eqnarray}
which generates the Reissner-Nordstrom black hole in Rainbow
gravity \cite{RN}. For $Q=0$, the above metric reduces to the
Schwarzschild black hole in Rainbow gravity \cite{Schw}. Also we
observe that the quantities $A(r),~B(r)$ depend on the Rastall
parameter $\lambda$ and Rainbow function $\Sigma(x)$.

\section{Physical Aspects of Parameters of Charged Gravastar}

Now we study the aspects of the physical parameters of the charged
gravastar like proper length of the shell, energy and entropy
within the shell. We shall also examine the impact of
electromagnetic field on different physical features of the
charged gravastar in the framework of Rastall-Rainbow gravity.

\subsection{Proper Length of the Thin Shell}

Since the radius of inner boundary of the shell of the gravastar
is $r=D$ and the radius of outer boundary of the shell is
$r=D+\epsilon$, where $\epsilon$ is the proper thickness of the
shell which is assumed to be very small (i.e., $\epsilon\ll 1$).
So the stiff perfect fluid propagates between two boundaries of
the thin shell region of the gravastar. Now, the proper length of
the shell is described by \cite{Mazur}
\begin{equation}
\ell=\int_{D}^{D+\epsilon}\sqrt{\frac{A(r)}{\Sigma^{2}(x)}}~dr
\end{equation}
Since in the shell region, the expressions of $A(r)$ is
complicated, so it is very difficult to obtain the analytical form
of the above integral. So let us assume
$\sqrt{A(r)}=\frac{dg(r)}{dr}$, so from the above integral we can
write
\begin{eqnarray}
&&\ell=\frac{1}{\Sigma(x)}\int_{D}^{D+\epsilon}\frac{dg(r)}{dr}~dr=\frac{1}{\Sigma(x)}[g(D+\epsilon)-g(D)]
\nonumber\\
&&\approx \left.\frac{\epsilon}{\Sigma(x)}
\frac{dg(r)}{dr}\right|_{D}=\frac{\epsilon\sqrt{A(D)}}{\Sigma(x)}
\end{eqnarray}
since $\epsilon\ll 1$, so $O(\epsilon^{2})\approx 0$. So in the
above manipulation, we have considered only the first order term
of $\epsilon$. Thus for this approximation, the proper length will
be
\begin{equation}
\ell=\frac{\epsilon}{\Sigma(x)}
\left[h_{1}+\frac{2(2\lambda-1)}{\lambda}~logD-\frac{E_{0}^{2}G}{\lambda(m+1)\Sigma^{2}(x)}~D^{2m+2}
\right]^{-\frac{1}{2}}
\end{equation}
The above result shows that the proper length of the thin shell of
the gravastar is proportional to the thickness $\epsilon$ of the
shell. We observe that proper length of the thin shell depends on
the electric field $E_{0}$ of the gravastar, Rastall parameter
$\lambda$ and Rainbow function $\Sigma(x)$. Due to Awad et al
\cite{Awad} and Khodadi et al \cite{Khod}, here we have chosen
$\Sigma(x)=\sqrt{1+x^{2}}$ where $x={\cal E}/{\cal E}_{Pl}$. We
have plotted the proper length $\ell$ vs thickness $\epsilon$ and
radius $D$ in fig. 1 and fig.2 respectively. From the figures, we
have seen that the proper length $\ell$ increases with the
thickness $\epsilon$ of the shell of the charged gravastar but
decreases as radius $D$ increases. On the other hand, we have also
plotted the proper length $\ell$ vs Rastall parameter $\lambda$
and test particle's charge ${\cal E}$ in fig. 3. From the figure,
we have observed that the proper length $\ell$ very slowly
decreases as Rastall parameter $\lambda$ increases and smoothly
decreases as the test particle's charge ${\cal E}$ increases.  \\

\subsection{Energy of the Charged Gravastar}

The energy content within the shell region of the charged
gravastar is given as \cite{Ghos}
\begin{eqnarray}
&&{\cal W}=\int_{D}^{D+\epsilon}4\pi r^{2}\bar{\rho}~dr \nonumber\\
&&=\frac{1}{2\lambda
G\Sigma^{2}(x)}\left[\frac{2GE_{0}^{2}}{2m+3}\left((D+\epsilon)^{2m+3}-D^{2m+3}
\right) \right. \nonumber\\
&&\left. -\epsilon(3\lambda-2)\Sigma^{2}(x) \right]
\end{eqnarray}
For the approximation $\epsilon\ll 1$ due to thin shell region of
the gravastar, we may obtain
\begin{eqnarray}
{\cal W}=\frac{\epsilon}{2\lambda
G\Sigma^{2}(x)}\left[2GE_{0}^{2}D^{2m+2}-(3\lambda-2)\Sigma^{2}(x)
\right]
\end{eqnarray}
We see that the energy content in the shell is proportional to the
thickness ($\epsilon$) of the shell. Also we observe that the
energy of the gravastar depends on the electric field $E_{0}$ of
the gravastar, Rastall parameter $\lambda$ and Rainbow function
$\Sigma(x)$. We have plotted the energy content ${\cal W}$ in the
shell vs thickness $\epsilon$ and radius $D$ in fig. 4 and fig.5
respectively. From the figures, we have seen that the the energy
content ${\cal W}$ in the shell increases with the thickness
$\epsilon$ of the shell of the charged gravastar as well as the
radius $D$. On the other hand, we have also plotted the energy
content ${\cal W}$ in the shell vs Rastall parameter $\lambda$ and
test particle's charge ${\cal E}$ in fig. 6. From the figure, we
have observed that the energy content ${\cal W}$ in the shell
decreases as increase of the Rastall parameter $\lambda$ and the
test particle's charge ${\cal E}$.  \\

\subsection{Entropy of the Charged Gravastar}

Entropy is the disorderness within the body of a gravastar. Mazur
and Mottola \cite{Mazur,Mazur1} have shown that the entropy
density in the interior region $R_{1}$ of the gravastar is zero.
But, the entropy within the thin shell can be described by
\cite{Mazur}
\begin{equation}
{\cal S}=\int_{D}^{D+\epsilon} 4\pi r^{2} s(r)
\sqrt{\frac{A(r)}{\Sigma^{2}(x)}}~dr
\end{equation}
where $s(r)$ is the entropy density corresponding to the specific
temperature $T(r)$ and which can be defined as
\begin{equation}
s(r)=\frac{\gamma^{2}k_{B}^{2}T(r)}{4\pi \hbar}=\frac{\gamma
k_{B}}{\hbar}\sqrt{\frac{p(r)}{2\pi}}
\end{equation}
where $\gamma$ is dimensionless constant. So entropy within the
thin shell can be written as
\begin{eqnarray}
&&{\cal S}=\frac{\gamma
k_{B}}{\hbar\Sigma(x)}\int_{D}^{D+\epsilon} r^{2}
\sqrt{8\pi p(r)~A(r)}~dr \nonumber\\
&&=\frac{\gamma k_{B}}{\hbar\sqrt{\lambda
G}~\Sigma(x)}\int_{D}^{D+\epsilon}r\left[GE_{0}^{2}r^{2m+2}+(1-2\lambda)
\Sigma^{2}(x)\right]^{\frac{1}{2}}\times \nonumber\\
&&\left[h_{1}+\frac{2(2\lambda-1)}{\lambda}~logr-\frac{E_{0}^{2}G}{\lambda(m+1)\Sigma^{2}(x)}~r^{2m+2}
\right]^{-\frac{1}{2}} dr \nonumber\\
\end{eqnarray}
Now it is very difficult to obtain analytical form of the above
integral. So using the approximation $\epsilon\ll 1$, we can
obtain
\begin{eqnarray}
&&{\cal S}\approx\frac{\epsilon\gamma k_{B}}{\hbar\Sigma(x)}~D^{2}
\sqrt{8\pi p(D)~A(D)}  \nonumber\\
&&\approx\frac{\epsilon\gamma k_{B}D}{\hbar\sqrt{\lambda
G}~\Sigma(x)}~\left[GE_{0}^{2}D^{2m+2}+(1-2\lambda)\Sigma^{2}(x)\right]^{\frac{1}{2}}\times \nonumber\\
&&\left[h_{1}+\frac{2(2\lambda-1)}{\lambda}~logD-\frac{E_{0}^{2}G}{\lambda(m+1)\Sigma^{2}(x)}~D^{2m+2}
\right]^{-\frac{1}{2}} \nonumber\\
\end{eqnarray}
This result shows that the entropy in the shell of the charged
gravastar is proportional to the thickness $\epsilon$ of the
shell. We observe that the entropy depends on the electric field
$E_{0}$ of the gravastar, Rastall parameter $\lambda$ and Rainbow
function $\Sigma(x)$. We have plotted the entropy ${\cal S}$
within the shell vs thickness $\epsilon$ and radius $D$ in fig. 7
and fig.8 respectively. From the figures, we have seen that the
the entropy ${\cal S}$ within the shell increases with the
thickness $\epsilon$ of the shell of the charged gravastar as well
as the radius $D$. On the other hand, we have also plotted the
entropy ${\cal S}$ within the shell vs Rastall parameter $\lambda$
and test particle's charge ${\cal E}$ in fig. 9. From the figure,
we have observed that the entropy ${\cal S}$ within the shell
decreases as increase of the Rastall parameter $\lambda$ and the
test particle's charge ${\cal E}$.  \\

\begin{figure}
\includegraphics[height=1.5in]{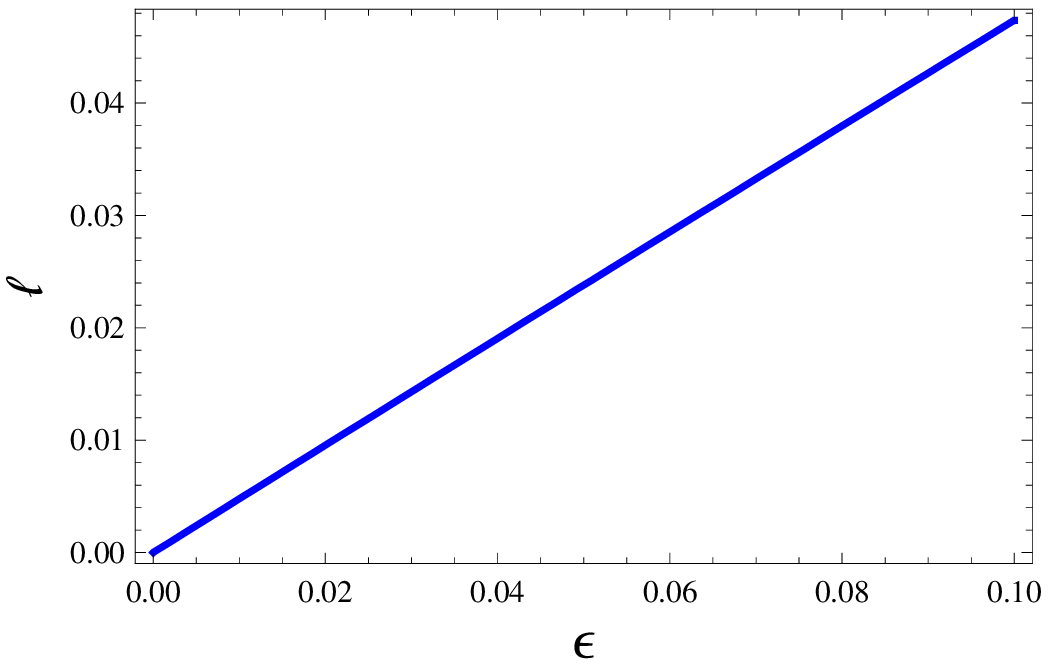}\\

\vspace{2 mm} Fig. 1 represents the plot proper length $\ell$ vs
thickness $\epsilon$.\\

\vspace{2 mm}

\includegraphics[height=1.5in]{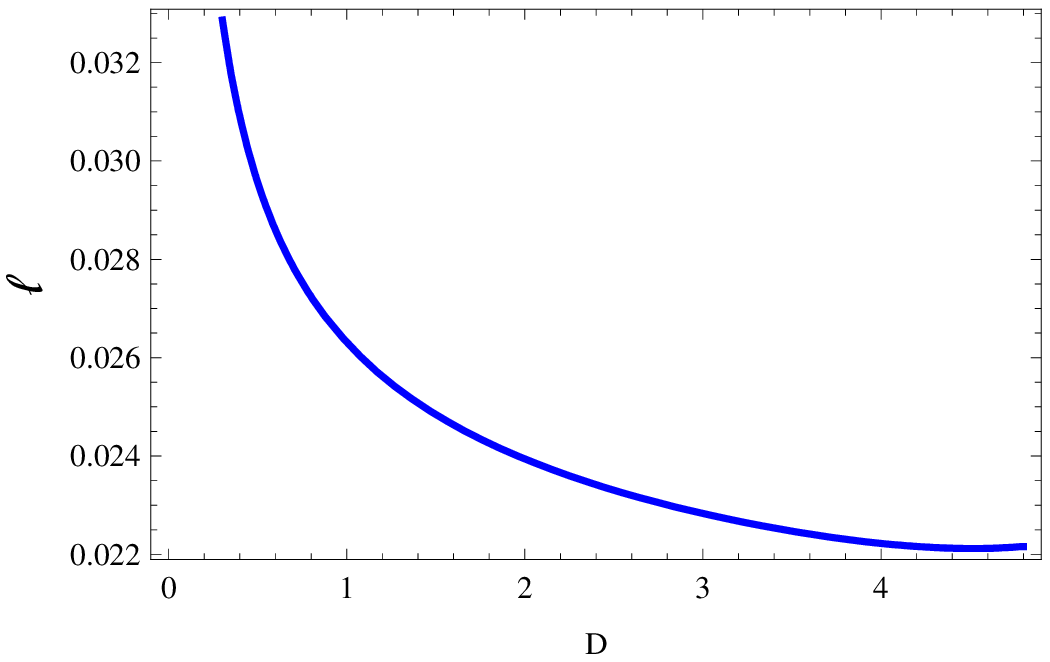}\\

\vspace{2 mm} Fig. 2 represents the plot of proper length $\ell$
vs radius $D$.

\vspace{2 mm}

\includegraphics[height=2.0in]{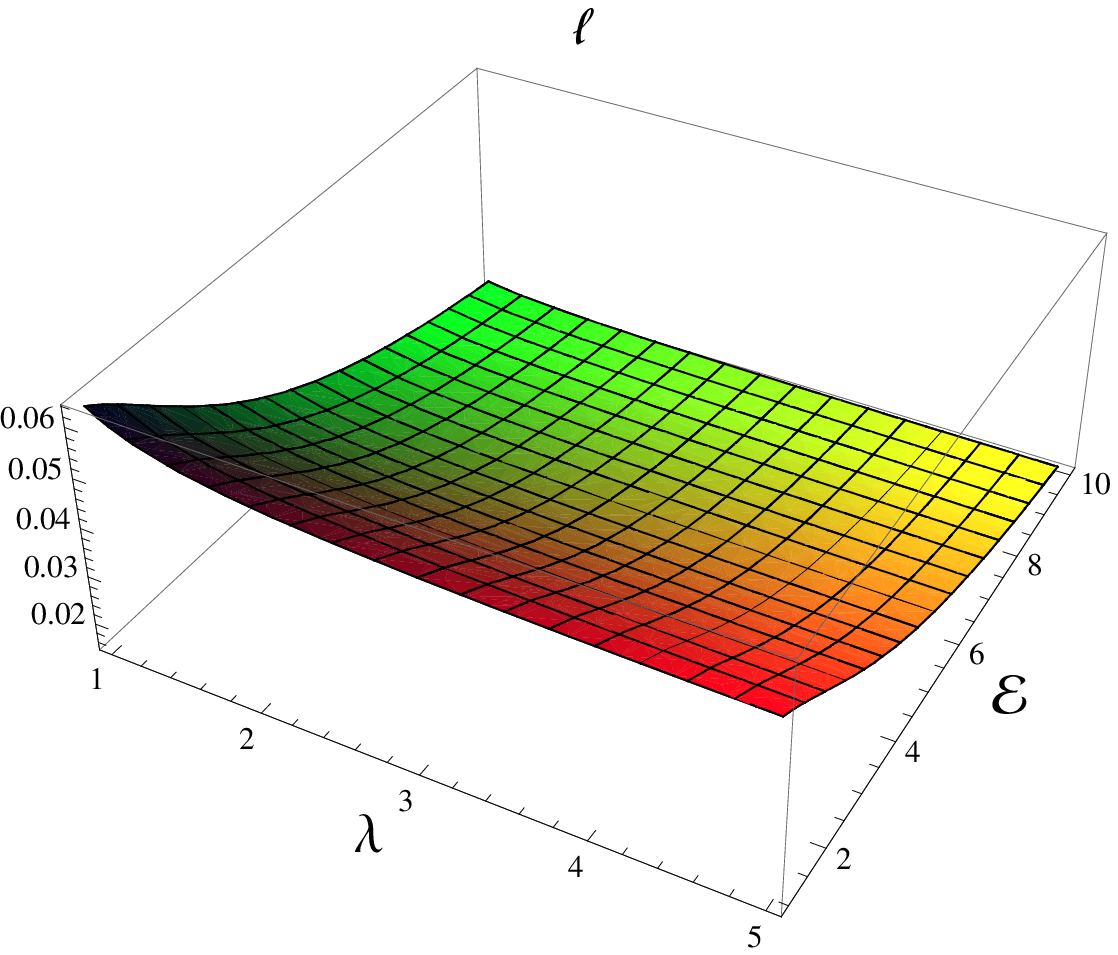}\\

\vspace{2 mm} Fig. 3 represents the plot of proper length $\ell$
vs Rastall parameter $\lambda$ and test particle's charge ${\cal
E}$.
\end{figure}

\begin{figure}
\includegraphics[height=1.5in]{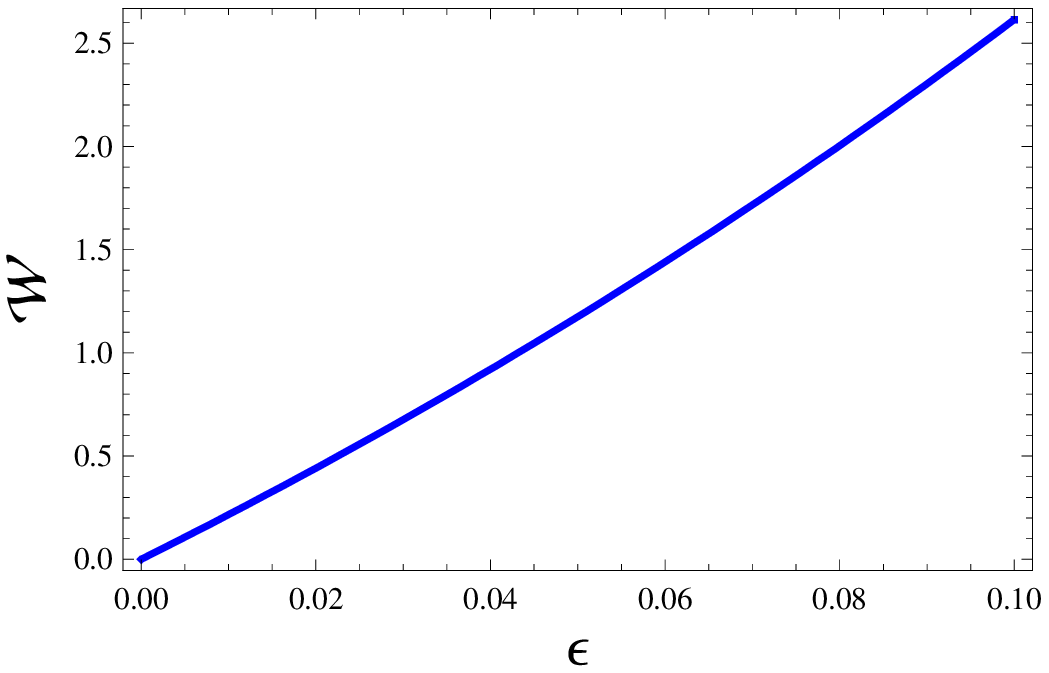}\\

\vspace{2 mm} Fig. 4 represents the plot of energy content ${\cal
W}$ vs thickness $\epsilon$.\\

\vspace{2 mm}

\includegraphics[height=1.5in]{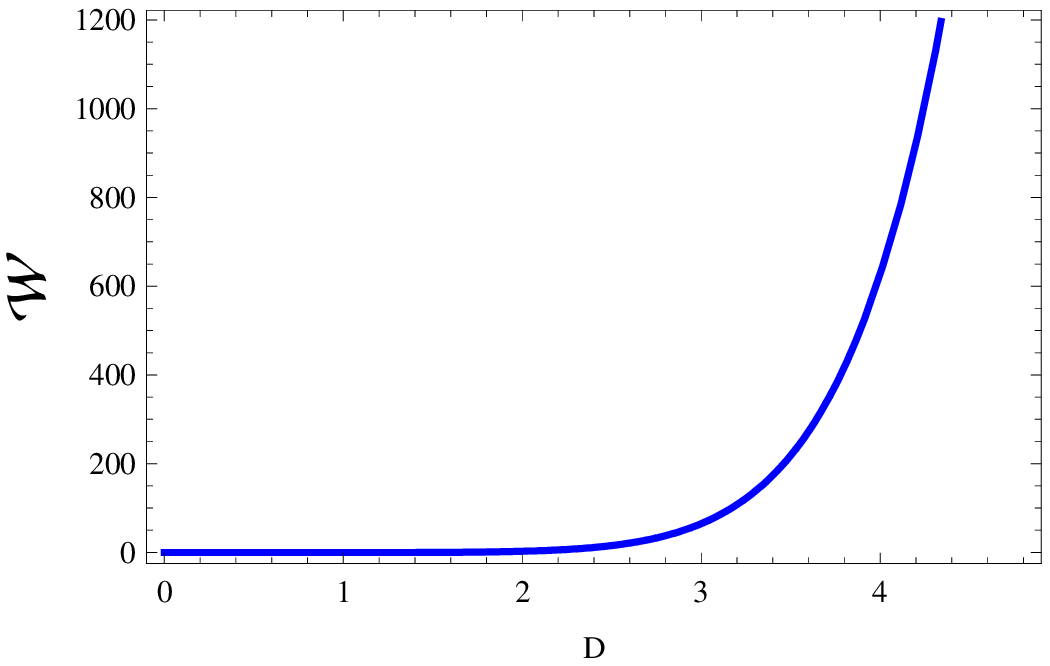}\\

\vspace{2 mm} Fig. 5 represents the plot of energy content ${\cal
W}$ vs radius $D$.

\vspace{2 mm}

\includegraphics[height=2.0in]{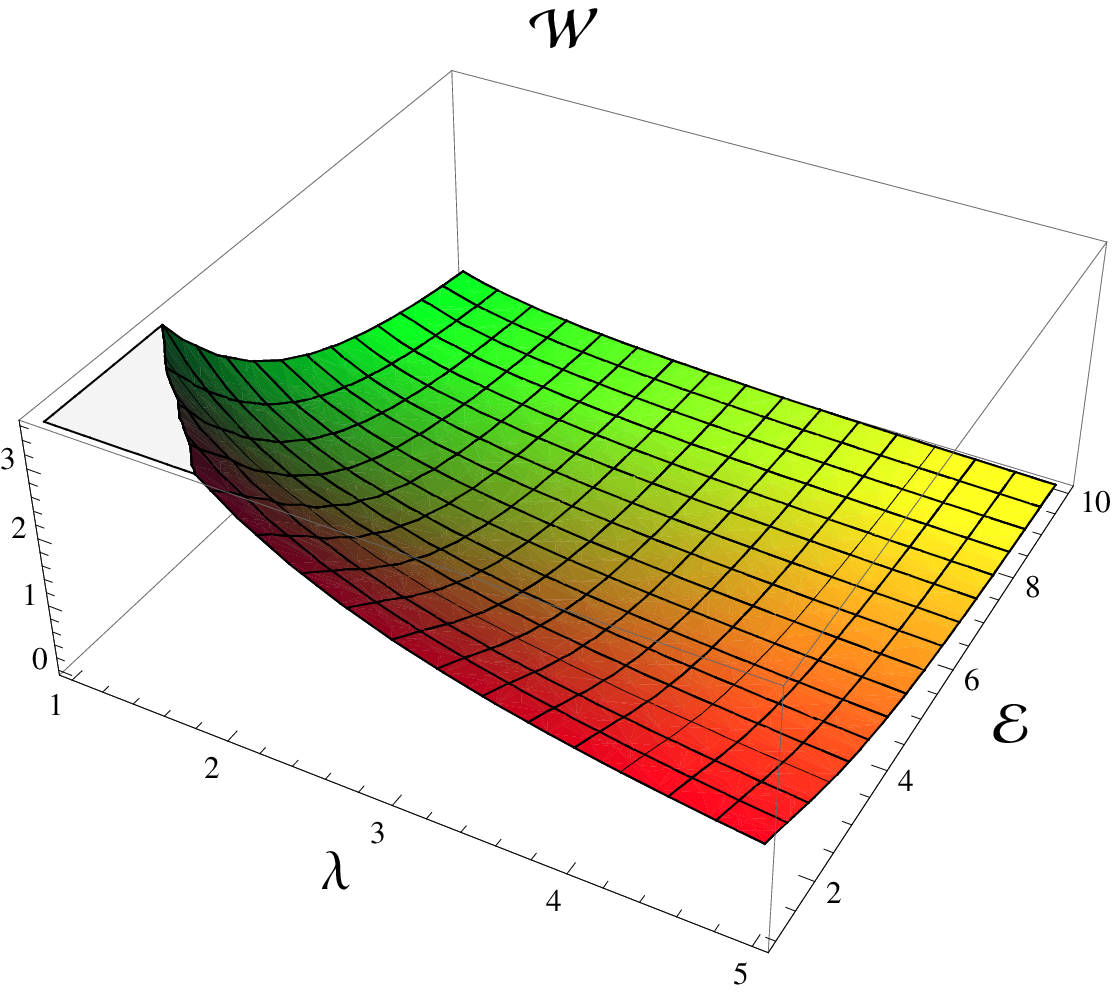}\\

\vspace{2 mm} Fig. 6 represents the plot of energy content ${\cal
W}$ vs Rastall parameter $\lambda$ and test particle's charge
${\cal E}$.
\end{figure}

\begin{figure}
\includegraphics[height=1.5in]{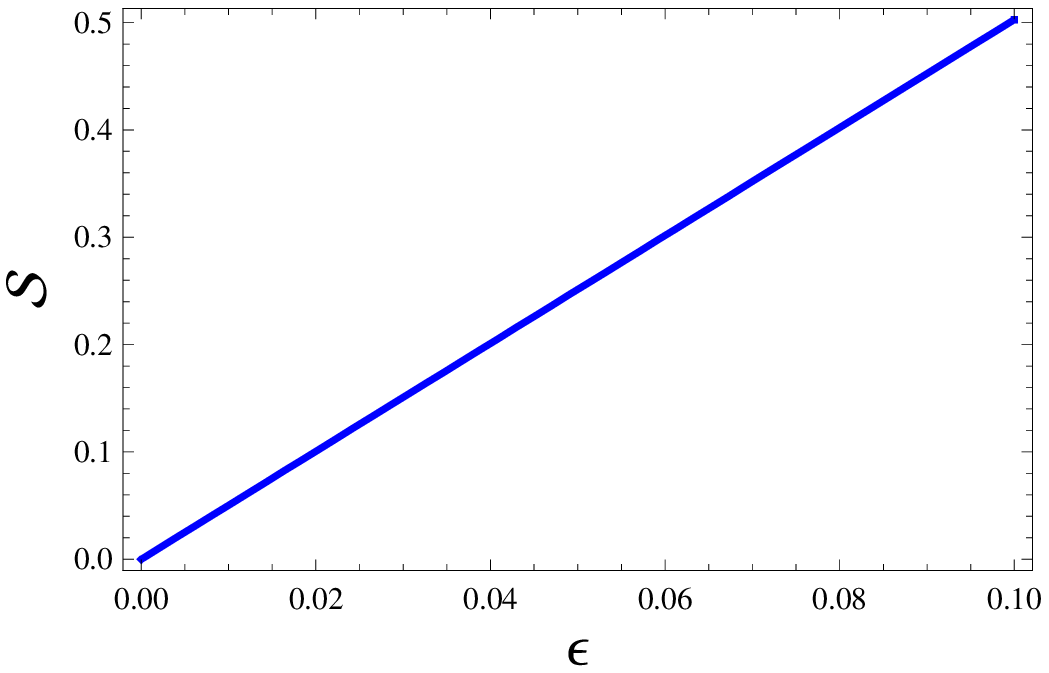}\\

\vspace{2 mm} Fig. 7 represents the plot of entropy ${\cal
S}$ vs thickness $\epsilon$.\\

\vspace{2 mm}

\includegraphics[height=1.5in]{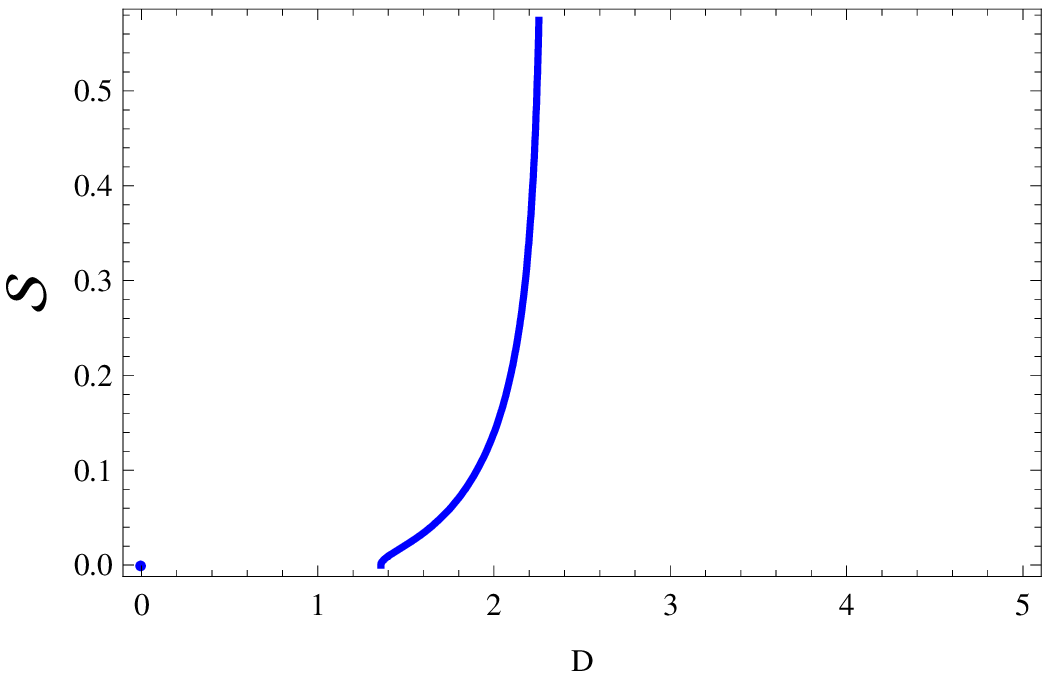}\\

\vspace{2 mm} Fig. 8 represents the plot of entropy ${\cal S}$ vs
radius $D$.

\vspace{2 mm}

\includegraphics[height=2.0in]{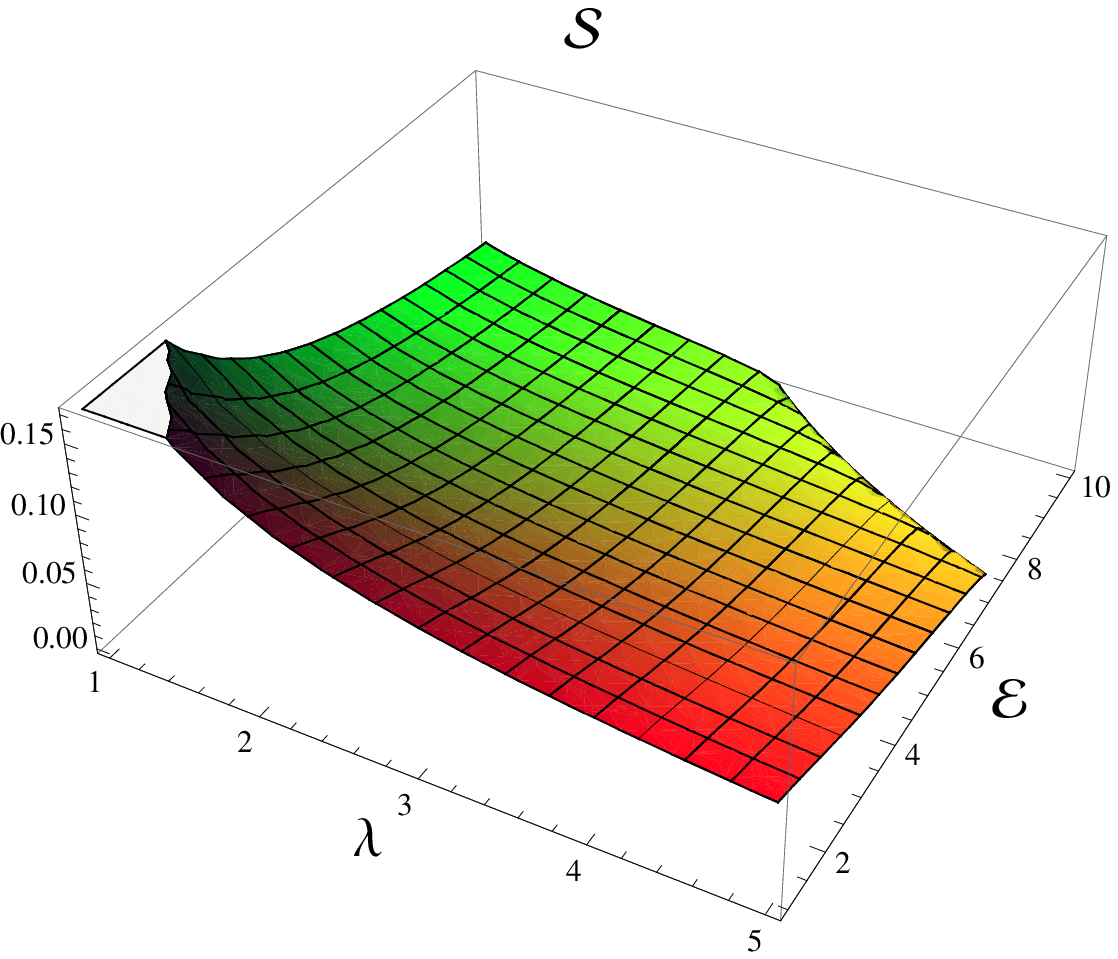}\\

\vspace{2 mm} Fig. 9 represents the plot of entropy ${\cal S}$ vs
Rastall parameter $\lambda$ and test particle's charge ${\cal E}$.
\end{figure}

\section{Junction Conditions between Interior and Exterior Regions}

Since gravastar consists of three regions: interior region, thin
shell region and exterior region, so according to the
Darmois-Israel formalism \cite{Is1,Is2,Is3} we want to study the
matching between the surfaces of the interior and exterior
regions. We denote $\sum$ is the junction surface which is located
at $r=D$. In the Rainbow gravity, we consider the metric on the
junction surface as in the form
\begin{equation}
ds^{2}=-\frac{f(r)}{\Pi^{2}(x)}dt^{2}+\frac{1}{\Sigma^{2}(x)f(r)}dr^{2}
+\frac{r^{2}}{\Sigma^{2}(x)}(d\theta^{2}+\sin^{2}\theta d\phi^{2})
\end{equation}
where the metric coefficients are continuous at the junction
surface $\sum$, but their derivatives might not be continuous at
$\sum$. In the joining surface $S$, the surface tension and
surface stress energy may be resolved by the discontinuity of the
extrinsic curvature of $S$ at $r = D$. The expression of the
stress-energy surface $S_{ij}$ is defined by the Lanczos equation
\cite{Lanc} (with the help of Darmois-Israel formalism)
\begin{equation}
S_{j}^{i}=-\frac{1}{8\pi}(\eta_{j}^{i}-\delta_{j}^{i}\eta_{k}^{k})
\end{equation}
where $\eta_{ij}=K_{ij}^{+}-K_{ij}^{-}$. Here $K_{ij}$ denotes the
extrinsic curvature. Here the signs $``-"$ and $``+"$ respectively
correspond to the interior and the exterior regions of the
gravastar. So $\eta_{ij}$ describes the discontinuous surfaces in
the second fundamental forms of the extrinsic curvatures. The
extrinsic curvatures on the both surfaces of the shell region can
be described by
\begin{equation}
K_{ij}^{\pm}=\left[-n_{\nu}^{\pm}\left\{\frac{\partial^{2}x_{\nu}}{\partial
\xi^{i}\partial \xi^{j}}+\Gamma_{\alpha\beta}^{\nu}\frac{\partial
x^{\alpha}}{\partial \xi^{i}}\frac{\partial x^{\beta}}{\partial
\xi^{j}} \right\}\right]_{\sum}
\end{equation}
where $\xi^{i}$ represent the intrinsic coordinates on the shell,
$n_{\nu}^{\pm}$ describe the unit normals to the surface $\sum$ of
the gravastar, defined by $n_{\nu}n^{\nu}=-1$. For the above
metric, we can obtain
\begin{equation}
n_{\nu}^{\pm}=\pm\left[g^{\alpha\beta}   \frac{\partial
f}{\partial x^{\alpha}} \frac{\partial f}{\partial
x^{\beta}}\right]^{-\frac{1}{2}} \frac{\partial f}{\partial
x^{\nu}}
\end{equation}
According to the Lanczos equation, the stress-energy surface
tensor can be written as $S^{i}_{j}=diag(-\varrho,{\cal P},{\cal
P},{\cal P})$ where $\varrho$ is the surface energy density and
${\cal P}$ is the surface pressure. Using the matching conditions,
in the charged gravastar model, the surface energy density and the
surface pressure can be obtained as

\begin{eqnarray}
&&\varrho(D)=\frac{1}{4\pi D}\left(1-q_{2}(x)D^{2}
+q_{3}(x)D^{2m+2} \right)^{\frac{1}{2}}
\nonumber\\
&& -\frac{1}{4\pi D}\left(1-\frac{2GM}{
D}+\frac{q_{1}(x)}{D^{2}}\right)^{\frac{1}{2}}
\end{eqnarray}
and
\begin{eqnarray}
&&{\cal P}(D)=\frac{1}{8\pi D}\left(1-\frac{2GM}{
D}+\frac{q_{1}(x)}{D^{2}}\right)^{\frac{1}{2}}
\nonumber\\
&&-\frac{1}{8\pi D}\left(1-q_{2}(x)D^{2} +q_{3}(x)D^{2m+2}
\right)^{\frac{1}{2}} \nonumber \\
&&+\frac{1}{16\pi} \left(\frac{GM}{ D^{2}}-\frac{q_{1}(x)}{
D^{3}}\right) \left(1-\frac{2GM}{
D}+\frac{q_{1}(x)}{D^{2}}\right)^{-\frac{1}{2}} \nonumber \\
&&-\frac{1}{16\pi} \left(-q_{2}(x)D +(m+1)q_{3}(x)D^{2m+1}
\right)\times \nonumber\\
&&\left(1-q_{2}(x)D^{2} +q_{3}(x)D^{2m+2} \right)^{-\frac{1}{2}}
\end{eqnarray}

where
\begin{eqnarray}
&&q_{1}(x)=\frac{GQ^{2}}{(2\lambda-1)\Sigma^{2}(x)},
q_{2}(x)=\frac{8\pi Gk_{2}}{3(2\lambda-1)\Sigma^{2}(x)},
\nonumber\\
&&q_{3}(x)=\frac{2G}{m(2m+3)(2\lambda-1)\Sigma^{2}(x)}.
\end{eqnarray}

We have drawn the plots of surface energy density $\varrho(D)$ and
surface pressure ${\cal P}(D)$ vs radius $D$ in fig. 10 and fig.
12. From these figures we have seen that the surface energy
density $\varrho(D)$ increases and surface pressure ${\cal P}(D)$
decreases as radius $D$ increases. On the other hand, we have
drawn the plots of surface energy density $\varrho(D)$ and surface
pressure ${\cal P}(D)$ vs Rastall parameter $\lambda$ and test
particle's charge ${\cal E}$ in fig. 11 and fig. 13. We have
observed that if Rastall parameter $\lambda$ and test particle's
charge ${\cal E}$ increase then surface energy density
$\varrho(D)$ decreases but surface pressure ${\cal P}(D)$
increases. Also we have seen that surface pressure ${\cal P}(D)$
is always keep negative sign.

\subsection{Equation of State}
The equation of state parameter $w(D)$ (at $r=D$) can be described
as
\begin{equation}
w(D)=\frac{{\cal P}(D)}{\varrho(D)}
\end{equation}
So the equation of state parameter can be obtained in the
following form
\begin{eqnarray}
&&w(D)=-\frac{1}{2}+\frac{1}{4} \left[ \left(\frac{GM}{
D^{2}}-\frac{q_{1}(x)}{ D^{3}}\right) \times \right. \nonumber\\
&&\left(1-\frac{2GM}{
D}+\frac{q_{1}(x)}{D^{2}}\right)^{-\frac{1}{2}} \nonumber \\
&&-\left(-q_{2}(x)D +(m+1)q_{3}(x)D^{2m+1}
\right)\times \nonumber\\
&&\left. \left(1-q_{2}(x)D^{2} +q_{3}(x)D^{2m+2}
\right)^{-\frac{1}{2}} \right]\times \nonumber\\
&&\left[\left(1-q_{2}(x)D^{2} +q_{3}(x)D^{2m+2}
\right)^{\frac{1}{2}} \right. \nonumber\\
&&\left. -\left(1-\frac{2GM}{
D}+\frac{q_{1}(x)}{D^{2}}\right)^{\frac{1}{2}} \right]^{-1}
\end{eqnarray}

We have plotted the equation of state parameter $w(D)$ vs radius
$D$ in fig. 14 and seen that $w(D)$ increases as radius $D$
increases. Also we have plotted the equation of state parameter
$w(D)$ vs Rastall parameter $\lambda$ and test particle's charge
${\cal E}$ in fig. 15 and it decreases as Rastall parameter
$\lambda$ and test particle's charge ${\cal E}$ increase but
$w(D)$ is always keep negative sign.

\subsection{Mass}

The mass ${\cal M}$ of the thin shell of the charged gravastar can
be obtained from the following formula
\begin{equation}
{\cal M}=4\pi D^{2}\varrho(D)
\end{equation}
which can be expressed as in the form
\begin{eqnarray}
&&{\cal M}=D\left[\left(1-q_{2}(x)D^{2} +q_{3}(x)D^{2m+2}
\right)^{\frac{1}{2}} \right. \nonumber\\
&& \left. -\left(1-\frac{2GM}{
D}+\frac{q_{1}(x)}{D^{2}}\right)^{\frac{1}{2}} \right]
\end{eqnarray}
So the total mass $M$ of the charged gravastar in terms of the
mass of the thin shell can be expressed as
\begin{eqnarray}
&&M=\frac{D}{2G}+\frac{q_{1}(x)}{2GD}
\nonumber\\
&&-\frac{D}{2G}\left[\left(1-q_{2}(x)D^{2} +q_{3}(x)D^{2m+2}
\right)^{\frac{1}{2}}~-\frac{{\cal M}}{D} \right]^{2}
\end{eqnarray}
We see that the total mass $M$ of the gravastar will be less than
$\frac{D^{2}+q_{1}(x)}{2GD}$. We have plotted the mass ${\cal M}$
of the thin shell of the charged gravastar vs radius $D$ in fig.
16 and seen that the mass ${\cal M}$ increases as radius $D$
increases. Also we have plotted the the mass ${\cal M}$ vs Rastall
parameter $\lambda$ and test particle's charge ${\cal E}$ in fig.
17 and it decreases as Rastall parameter $\lambda$ and test
particle's charge ${\cal E}$ increase.

\begin{figure}
\includegraphics[height=1.5in]{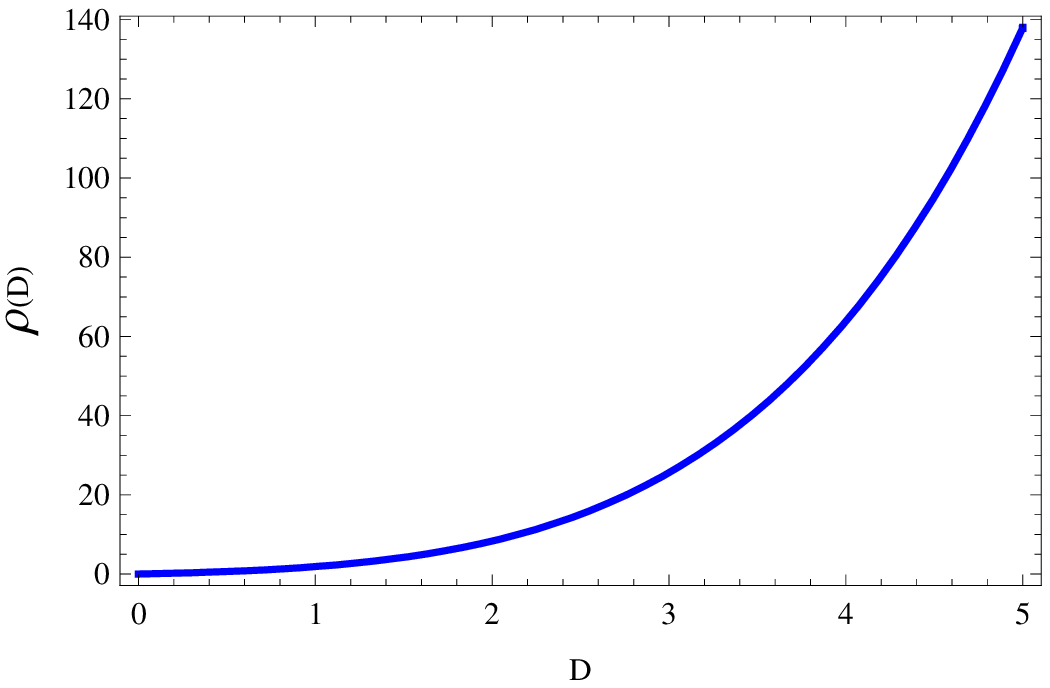}\\

\vspace{2 mm} Fig. 10 represents the plot of surface energy
density $\varrho(D)$ vs radius $D$.\\

\vspace{2 mm}

\includegraphics[height=2.0in]{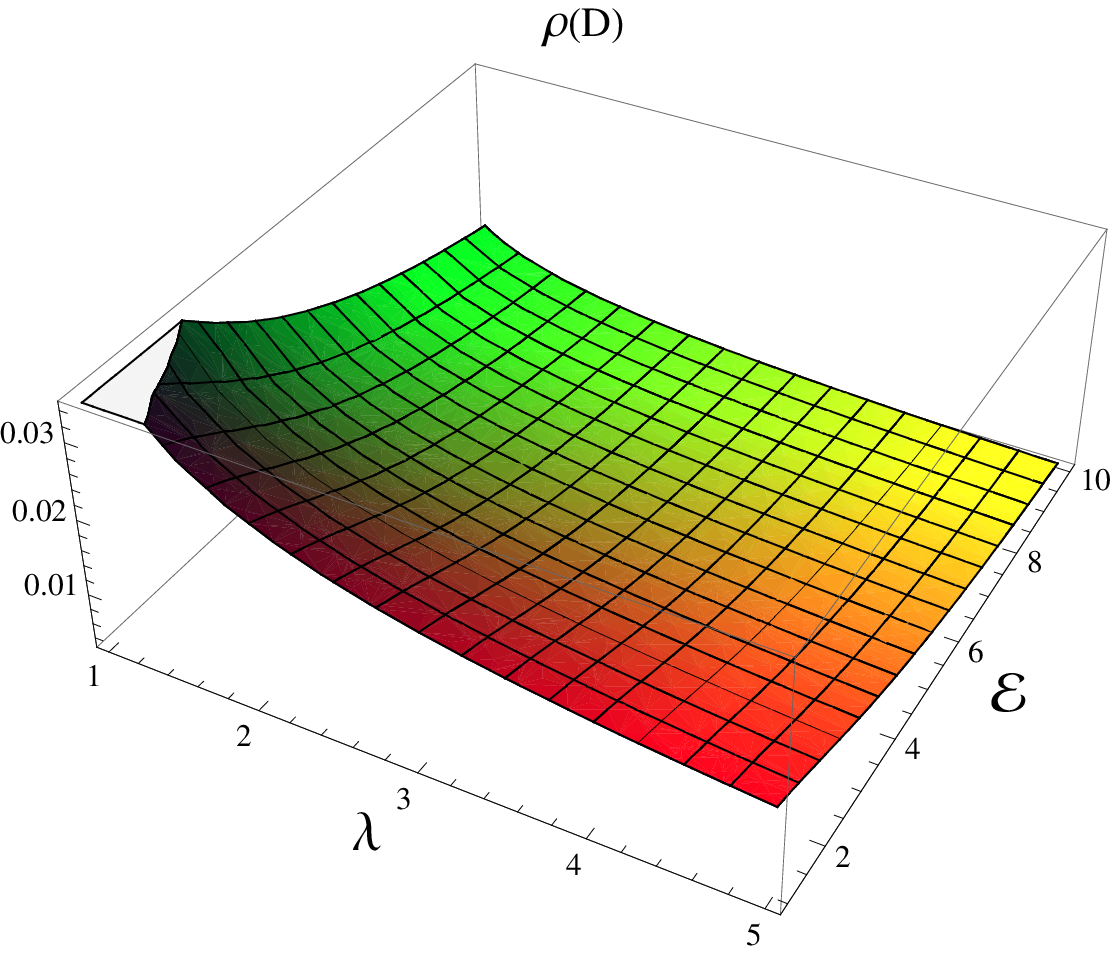}\\

\vspace{2 mm} Fig. 11 represents the plot of surface energy
density $\varrho(D)$ vs Rastall parameter $\lambda$ and test
particle's charge ${\cal E}$.
\end{figure}

\begin{figure}
\includegraphics[height=1.5in]{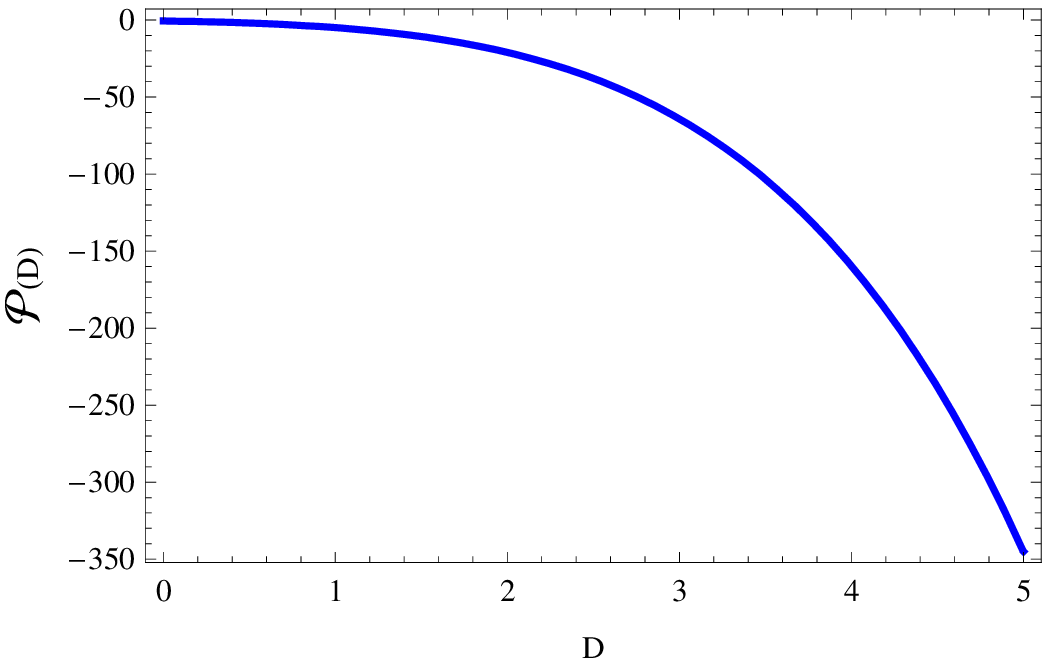}\\

\vspace{2 mm} Fig. 12 represents the plot of surface pressure ${\cal P}(D)$ vs radius $D$.\\

\vspace{2 mm}

\includegraphics[height=2.0in]{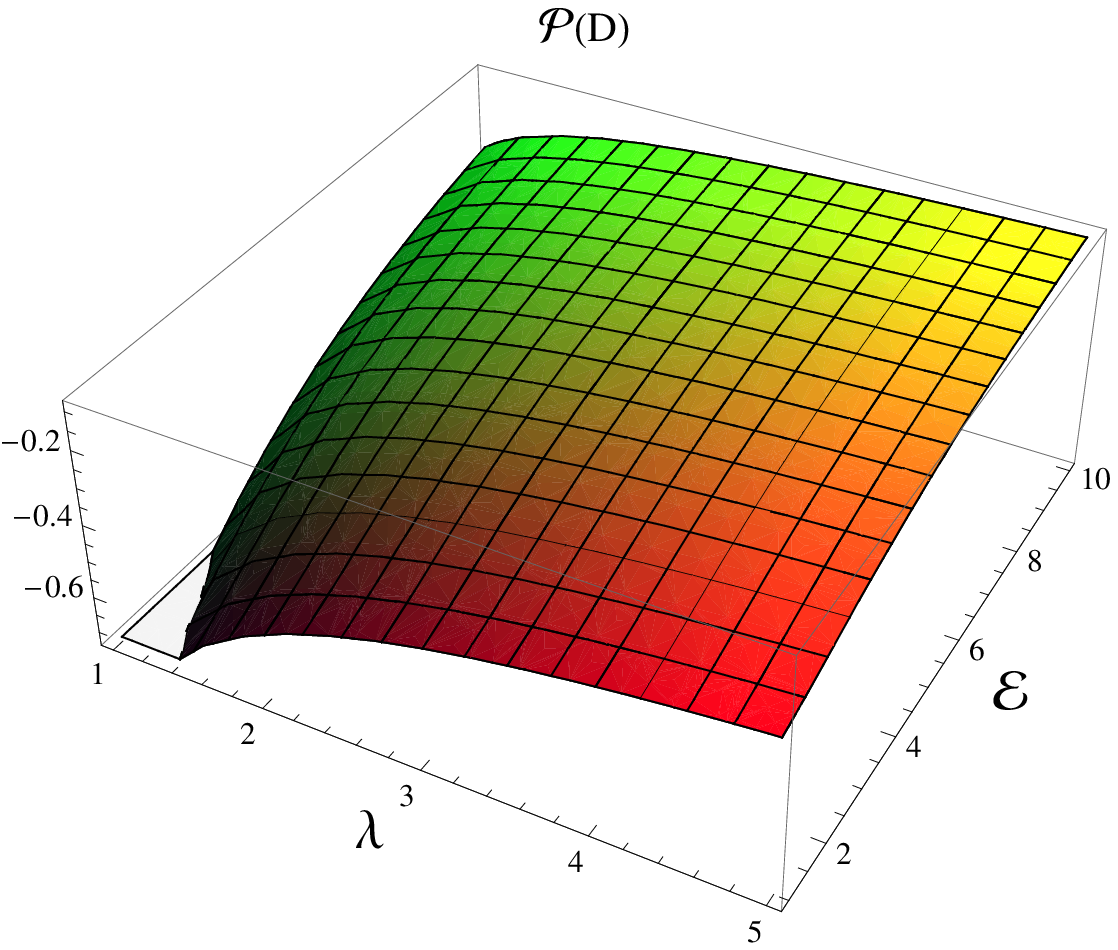}\\

\vspace{2 mm} Fig. 13 represents the plot of surface pressure
${\cal P}(D)$ vs Rastall parameter $\lambda$ and test particle's
charge ${\cal E}$.
\end{figure}

\begin{figure}
\includegraphics[height=1.5in]{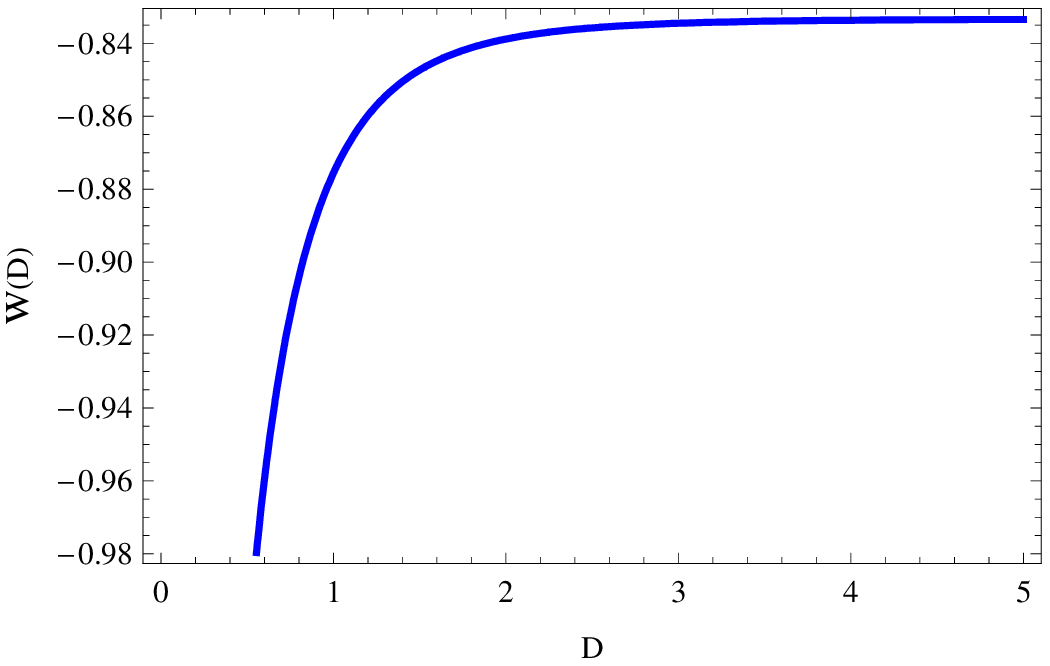}\\

\vspace{2 mm} Fig. 14 represents the plot of equation of state parameter $w(D)$ vs radius $D$.\\

\vspace{2 mm}

\includegraphics[height=2.0in]{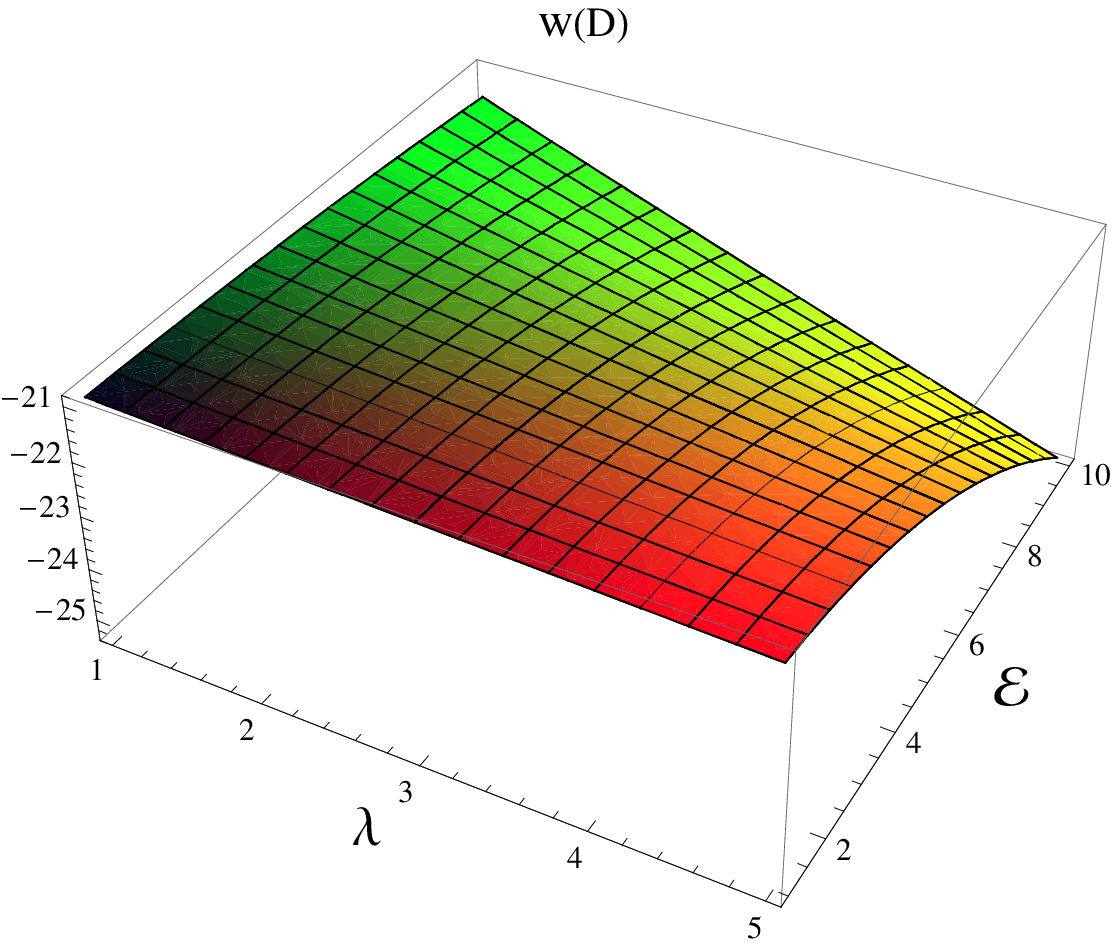}\\

\vspace{2 mm} Fig. 15 represents the plot of equation of state
parameter $w(D)$ vs Rastall parameter $\lambda$ and test
particle's charge ${\cal E}$.
\end{figure}

\begin{figure}
\includegraphics[height=1.5in]{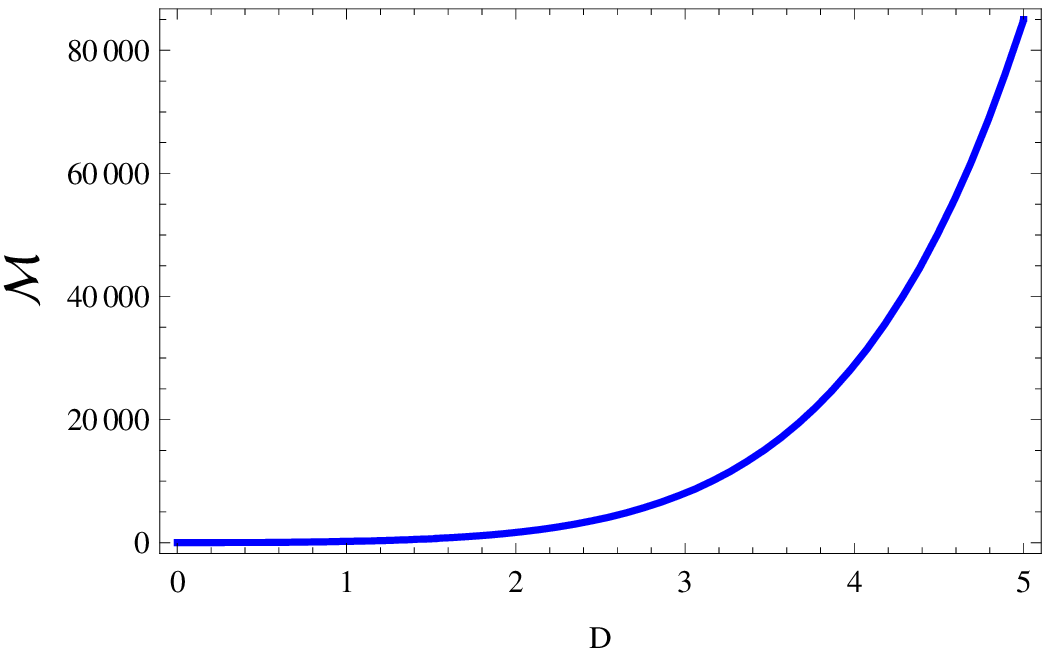}\\

\vspace{2 mm} Fig. 16 represents the plot of the mass of the thin shell ${\cal M}$ vs radius $D$.\\

\vspace{2 mm}

\includegraphics[height=2.0in]{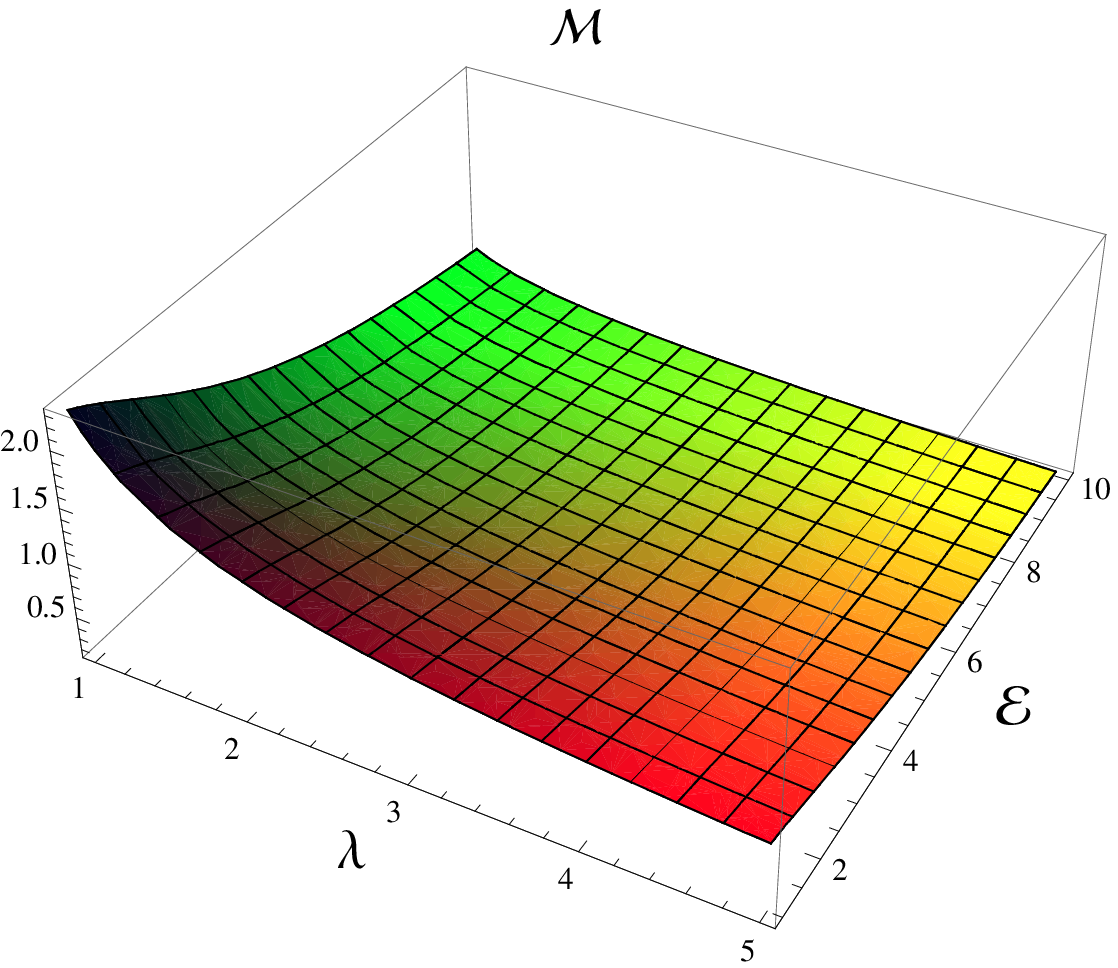}\\

\vspace{2 mm} Fig. 17 represents the plot of the mass of the thin
shell ${\cal M}$ vs Rastall parameter $\lambda$ and test
particle's charge ${\cal E}$.
\end{figure}

\subsection{Stability}

Poisson et al \cite{Poisson} have examined the linearized
stability for thin shell wormhole. Lobo et al \cite{Lobo} have
examined the linearized stability for thin shell wormhole with
cosmological constant. Ovgun et al \cite{Ali} have examined the
stability of charged thin-shell gravastar. Also Yousaf et al
\cite{Yous} have discussed the stability of the gravastar in
$f(R,{\cal T})$ gravity. Motivated by their work, we want to
analyze the stability of the charged gravastar in Rastall-Rainbow
gravity. For this purpose, define a parameter $\eta$ as follows
\cite{Poisson}:
\begin{equation}
\eta(D)=\frac{{\cal P}~'(D)}{\varrho~'(D)}
\end{equation}
The parameter $\eta$ is interpreted as the squared speed of sound
satisfying $0\le\eta\le 1$ normally. But for the stability of the
gravastar, the restriction may not be satisfied on the surface
layer. We have plotted $\eta$ vs radius $D$ in fig. 18 and seen
that $\eta$ decreases as radius $D$ increases. Also we have
plotted $\eta$ vs Rastall parameter $\lambda$ and test particle's
charge ${\cal E}$ in fig. 19 and it decreases as Rastall parameter
$\lambda$ and test particle's charge ${\cal E}$ increase. Figures
show that $\eta$ always keep positive sign which allows the
stability of the gravastar model.

\begin{figure}
\includegraphics[height=1.5in]{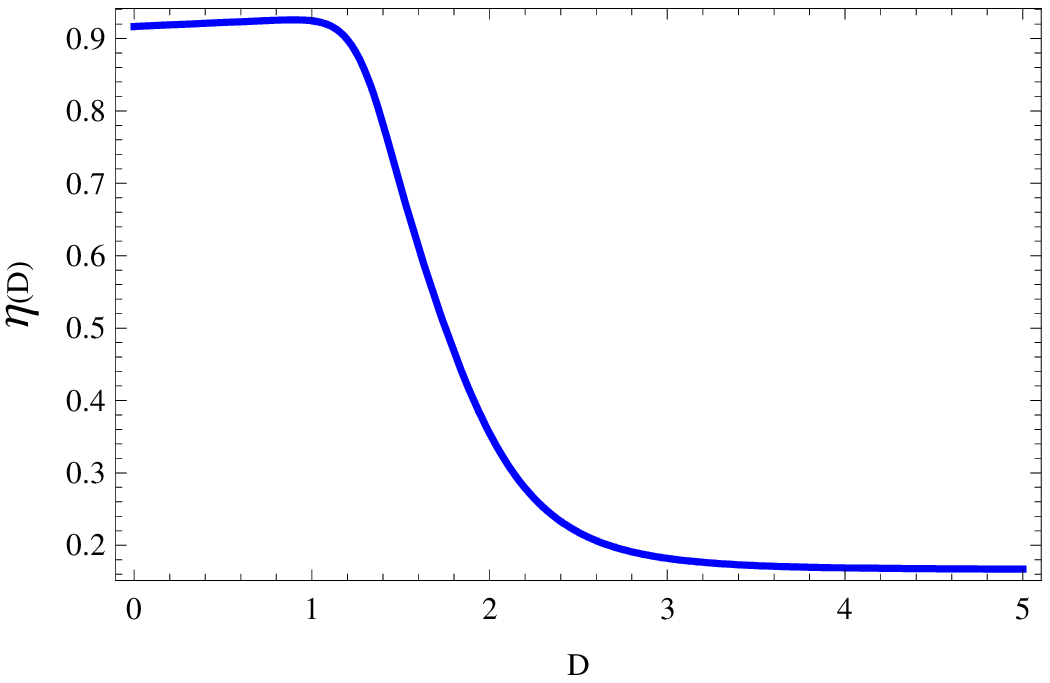}\\

\vspace{2 mm} Fig. 18 represents the plot of $\eta(D)$ vs radius $D$.\\

\vspace{2 mm}

\includegraphics[height=2.0in]{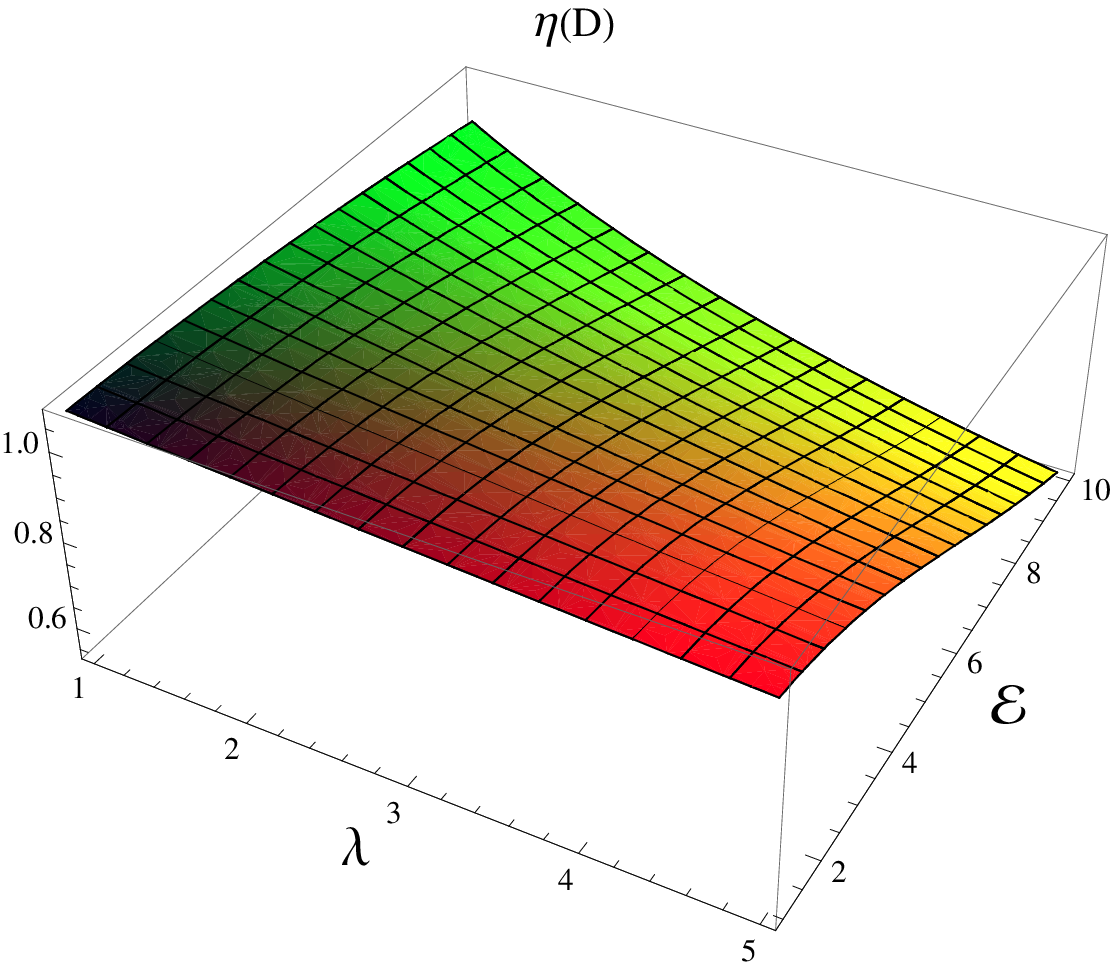}\\

\vspace{2 mm} Fig. 19 represents the plot of $\eta(D)$ vs Rastall
parameter $\lambda$ and test particle's charge ${\cal E}$.
\end{figure}

\section{Discussions}

In this work, we have considered the spherically symmetric stellar
system in the contexts of Rastall-Rainbow gravity theory in
presence of isotropic fluid source with electro-magnetic field.
The Einstein-Maxwell's field equations have been written in the
framework of Rastall-Rainbow gravity. Next we have discussed the
geometry of charged gravastar model. The gravastar consists of
three regions: interior region, thin shell region and exterior
region. In the interior region, the gravastar follows the equation
of sate (EoS) $p=-\rho$ and we have found the solutions of all
physical quantities like energy density, pressure, electric field,
charge density, gravitational mass and metric coefficients. In the
exterior region, we have obtained the exterior Riessner-Nordstrom
solution for vacuum model ($p=\rho=0$). In the shell region, the
fluid source follows the EoS $p=\rho$ (ultra-stiff perfect fluid).
In this region, since we have assumed that the interior and
exterior regions join together at a place, so the intermediate
region must be thin shell and the thickness of the shell of the
gravastar is infinitesimal. So the thin shell follows with the
limit $h~(\equiv A^{-1})\ll 1$. Under this approximation, we have
found the analytical solutions within the thin shell of the
gravastar. The physical quantities like the proper length of the
thin shell, energy content and entropy inside the thin shell of
the charged gravastar have been computed and we have shown that
they are directly proportional to the proper thickness of the
shell ($\epsilon$) due to the approximation ($\epsilon\ll 1$).
These physical quantities significantly depend on the Rastall
parameter $\lambda$ and Rainbow function $\Sigma(x)$ (which
depends on energy of the test particle). From figures 1 - 9, we
have seen that the proper length of the thin shell, energy content
and entropy inside the thin shell of the charged gravastar
increase as thickness increases. Also the proper length decreases
as radius increases. The energy content and entropy inside the
thin shell always increase as radius increases. Moreover, if
Rastall parameter $\lambda$ and test particle's charge ${\cal E}$
both increase then the proper length of the thin shell, energy
content and entropy inside the thin shell all decrease. According
to the Darmois-Israel formalism, we have studied the matching
between the surfaces of interior and exterior regions of the
charged gravastar and using the matching conditions, the surface
energy density and the surface pressure have been obtained. Also
the equation of state parameter on the surface, mass of the thin
shell have been obtained and the total mass of the charged
gravastar have been expressed in terms of the thin shell mass.
From figures 10, 14, 16 we have observed that the surface density,
equation of state and mass of the thin shell increase as radius
increases but from figure 12, we have seen that the surface
pressure always decreases as radius increases but keeps negative
sign and hence the equation of state keeps negative sign. Also the
figures 11, 15, 17 show that the surface density, equation of
state and mass of the thin shell decrease as the increase of both
Rastall parameter $\lambda$ and test particle's charge ${\cal E}$.
But figure 13 shows the surface pressure increases as the increase
of both Rastall parameter $\lambda$ and test particle's charge
${\cal E}$. Finally, we have explored the stable regions of the
charged gravastar in Rastall-Rainbow gravity in figures 18 and 19.
The figures show that $\eta$ always decreases but keeps in positive
sign as the increase of radius, Rastall parameter $\lambda$ and test
particle's charge ${\cal E}$.\\


\begin{thebibliography}{41}

\bibitem{Mazur} P. Mazur and E. Mottola, Report number: LA-UR-01-
5067 (arXiv:gr-qc/0109035).
\bibitem{Mazur1} P. Mazur and E. Mottola, Proc. Natl. Acad. Sci. (USA)
101, 9545 (2004).
\bibitem{Visser} M. Visser and D. L. Wiltshire, Class. Quant. Grav. 21, 1135
(2004).
\bibitem{De} A. DeBenedictis, D. Horvat, S. Ilijic, S. Kloster and K. S.
Viswanathan, Class. Quant. Grav. 23, 2303 (2006).
\bibitem{Cattoen} C. Cattoen, T. Faber and M. Visser, Class.
Quant. Grav. 22, 4189 (2005).
\bibitem{Bilic} N. Bilic, G. B. Tupper and R. D. Viollier, JCAP
0602, 013 (2006).
\bibitem{Carter} B. M. N Carter, Class. Quant. Grav. 22, 4551 (2005).
\bibitem{Usmani} A. A. Usmani, F. Rahaman, S. A. Rakib, S. Ray, K. K. Nandi, P. K. F. Kuhfittig and Z.
Hasan, Phys. Lett. B 701, 388 (2011).
\bibitem{Bhar} P. Bhar, Astrophys. Space Sci. 354, 2109 (2014).
\bibitem{Ayan} A. Banerjee, F. Rahaman, S. Islam and M. Govender, Eur. Phys. J. C 76, 34 (2016).
\bibitem{Farook1} F. Rahaman, S. Chakraborty, S.
Ray, A. A. Usmani and S. Islam, Int. J. Theor. Phys. 54, 50
(2015).
\bibitem{Ghos} S. Ghosh, F. Rahaman, B. K. Guha and S. Ray, Phys. Lett. B 767, 380
(2017).
\bibitem{Ghosh1} S. Ghosh, S. Ray, F. Rahaman and B. K. Guha, Annals of Phys. 394, 230
(2018).
\bibitem{Bene} B. M. N. Carter, Class. Quant. Grav. 22, 4551
(2005).
\bibitem{Rocha1} P. Rocha, R. Chan, M. F. A. da Silva and A. Wang, JCAP 0811, 010
(2008).
\bibitem{Rocha2} P. Rocha, A. Y. Miguelote, R. Chan, M. F. da Silva, N. O.
Santos and A. Wang, JCAP 0806, 025 (2008).
\bibitem{Chan1} R. Chan, M. F. A. da Silva, P. Rocha and A. Wang, JCAP 0903, 010
(2009).
\bibitem{R1} F. Rahaman, S. Ray, A. A. Usmani, Islam and S. Islam, Phys.
Lett. B 707, 319 (2012).
\bibitem{Lobo00} F. S. N. Lobo and R. Garattini, JHEP 1312, 065 (2013).
\bibitem{Ayan1} A. Banerjee, J. R. Villanueva, P. Channuie and K. Jusuf, Chin. Phys.
C 42, 115101 (2018).
\bibitem{Fel} F. de Felice, Y. Yu and J. Fang, Mon. Not. R. Astron. Soc.
277, L17 (1995).
\bibitem{Tur} B. V. Turimov, B. J. Ahmedov and A. A. Abdujabbarov, Mod.
Phys. Lett. A 24, 733 (2009).



\bibitem{Horvat} D. Horvat, S. Ilijic and A. Marunovic, Class. Quant.
Grav. 26, 025003 (2009).
\bibitem{Chan} R. Chan and M. F. A. da Silva, JCAP 1007, 029
(2010).
\bibitem{Rahaman1} F. Rahaman, A. A. Usmani, S. Ray and S. Islam, Phys. Lett. B 717, 1 (2012).
\bibitem{Brandt} C. F. C. Brandt, R. Chan, M. F. A. da Silva and P. Rocha, J. Mod. Phys. 6, 879 (2013).
\bibitem{Yous00} Z. Yousaf and M. Z. Bhatti, Mon. Not. R. Astron. Soc. 458,
1785 (2016).
\bibitem{Ali} A. Ovgun, A. Banerjee and K. Jusufi, Eur. Phys. J. C
77, 566 (2017).
\bibitem{Ray} S. Ray and B. Das, Gravit. Cosmol. 13, 224 (2007).

\bibitem{Boem} C. G. Boehmer, A. Mussa and N. Tamanini, Class. Quant. Grav.
28, 245020 (2011).
\bibitem{Del} C. Deliduman and B. Yapiskan, arXiv:1103.2225 [gr-qc].
\bibitem{Abha1} G. Abbas, S. Qaisar and M. A. Meraj, Astrophys. Space Sci. 357, 156 (2015).
\bibitem{Saha1} P. Saha and U. Debnath, Adv. High Energy Phys. 2018, 3901790
(2018).
\bibitem{Kp} A. V. Kpadonou, M. J. S. Houndj and M. E. Rodrigues, Astrophys. Space Sci. 361, 244 (2016).
\bibitem{Abha2} G. Abbas, S. Qaisar, W. Javed and M.A. Meraj, Iran. J. Sci. Technol. A 42, 1659 (2016).
\bibitem{Gan} M. G. Ganiou, C. Aïnamon, M. J. S. Houndjo and J. Tossa, Eur. Phys. J. Plus 132, 250 (2017).
\bibitem{Krori} K. D. Krori and J. Barua, J. Phys. A: Math. Gen. 8, 508 (1975).
\bibitem{Abbas6} G. Abbas, S. Nazeer and M. A. Meraj, Astrophys. Space Sci. 354, 449 (2014).
\bibitem{Abbas7} G. Abbas, A. Kanwal and M. Zubair, Astrophys. Space Sci. 357, 109 (2015).
\bibitem{Abbas8} G. Abbas et al, Astrophys. Space Sci. 357, 158 (2015).
\bibitem{Abbas9} G. Abbas, M. Zubair and G. Mustafa, Astrophys. Space Sci. 358, 26 (2015).
\bibitem{Das1} A. Das, S. Ghosh, B. K. Guha, S. Das, F. Rahaman and S. Ray,
Phys. Rev. D 95, 124011 (2017).
\bibitem{Yous} Z. Yousaf, K. Bamba, M. Z. Bhatti and U. Ghafoor, Phys. Rev. D 100, 024062
(2019).
\bibitem{Shamir} M. F. Shamir and M. Ahmad, Phys. Rev. D 97, 104031 (2018).


\bibitem{Rastall} P. Rastall, Phys. Rev. D 6, 3357 (1972).
\bibitem{Oliv} A. M. Oliveira, H. E. S. Velten, J. C. Fabris and L. Casarini, Phys. Rev. D 92, 044020 (2015).
\bibitem{AbhasS} G. Abbas and M. R. Shahzad, Eur. Phys. J. A 54, 211 (2018).
\bibitem{AbhasS1} G. Abbas and M. R. Shahzad, Astrophys. Space Sci.
363, 251 (2018).
\bibitem{AbhasS2} G. Abbas and M. R. Shahzad, Astrophys. Space Sci. 364, 50 (2019).

\bibitem{Mag} J. Magueijo and L. Smolin, Class. Quantum Grav. 21, 1725
(2004).
\bibitem{Hen} S. H. Hendi, G. H. Bordbar, B. E. Panah, S. Panahiyan, JCAP 1609, 013 (2016).
\bibitem{Gar} R. Garattini and G. Mandanici., arXiv:1601.00879
[physics.gen-ph].
\bibitem{Gara} R. Garattini and G. Mandanici, Eur. Phys. J. C 77, 57 (2017).
\bibitem{Mota0} C. E. Mota, L. C. N. Santos, G. Grams, F. M. da Silva and D. P. Menezes, Phys. Rev. D
100, 024043 (2019).
\bibitem{Debn} U. Debnath, Eur. Phys. J. C 79, 499 (2019).
\bibitem{Awad} A. Awad, A. F. Ali and B. Majumder, JCAP 1310, 052 (2013).
\bibitem{Khod} M. Khodadi, K. Nozari and H. R. Sepangi, Gen. Relativ. Grav.
48, 166 (2016).
\bibitem{Am} G. Amelino-Camelia et al, Nature 393, 763 (1998).




\bibitem{RN} A. Haldar and R. Biswas, Gen. Rel. Grav. 51, 72 (2019).
\bibitem{Schw} C. Leiva, J. Saavedra and J. Villanueva, Mod. Phys. Lett. A 24, 1443 (2009).



\bibitem{Is1} G. Darmois, Memorial des sciences mathematiques XXV,
Fasticule XXV, (Gauthier-Villars, Paris, France, 1927), chap. V.
\bibitem{Is2} W. Israel, Nuovo Cimemto 44, 1 (1966)
\bibitem{Is3} W. Israel, Nuovo Cimemto 48, 463(E) (1967).
\bibitem{Lanc} K. Lanczos, Ann. Phys. 379, 518 (1924).


\bibitem{Poisson} E. Poisson and M. Visser, Phys. Rev. D 52, 7318 (1995).
\bibitem{Lobo} F. S. N. Lobo and P. Crawford, Class. Quantum Gravity 21, 391
(2004).














\end{thebibliography}
\end{document}